\documentclass[twocolumn]{emulateapj}

\usepackage{makecell}
\usepackage{amsmath}
\usepackage[utf8]{inputenc}
\usepackage{graphicx}
\usepackage{color}
\usepackage[colorlinks]{hyperref}
\hypersetup{
    colorlinks,	
    citecolor=blue,
}

\newcommand{\be}{\begin{eqnarray}}
\newcommand{\ee}{\end{eqnarray}}

\newcommand{\src}{GX~339--4}
\newcommand{\hxmt}{\textit{Insight}--HXMT}

\shorttitle{Hard to soft transition of GX~339--4}
\shortauthors{Liu et al.}

\begin{document}

\title{The hard to soft transition of GX~339--4 as seen by \textit{Insight}--HXMT}

\author{Honghui Liu\altaffilmark{1}, Cosimo Bambi\altaffilmark{1, \dag}, Jiachen Jiang\altaffilmark{2}, Javier A. Garc{\'\i}a\altaffilmark{3,4}, Long Ji\altaffilmark{5}, Lingda Kong\altaffilmark{6}, Xiaoqin Ren\altaffilmark{7,8}, Shu Zhang\altaffilmark{7,8}, Shuangnan Zhang\altaffilmark{7,8}}

\altaffiltext{1}{Center for Field Theory and Particle Physics and Department of Physics, 
Fudan University, 200438 Shanghai, China. \email[\dag E-mail: ]{bambi@fudan.edu.cn}} 
\altaffiltext{2}{Institute of Astronomy, University of Cambridge, Madingley Road, Cambridge CB3 0HA, UK} 
\altaffiltext{3}{Cahill Center for Astronomy and Astrophysics, California Institute of Technology, 1216 E. California Boulevard, Pasadena, CA 91125, USA}
\altaffiltext{4}{Dr. Karl Remeis-Observatory and Erlangen Centre for Astroparticle Physics, Sternwartstr. 7, D-96049 Bamberg, Germany}
\altaffiltext{5}{School of Physics and Astronomy, Sun Yat-Sen University, 519082 Zhuhai, China}
\altaffiltext{6}{Institut f{\"u}r Astronomie und Astrophysik, Kepler Center for Astro and Particle Physics, Eberhard Karls, Universit{\"a}t, Sand 1, D-72076 T{\"u}bingen, Germany}
\altaffiltext{7}{Key Laboratory for Particle Astrophysics, Institute of High Energy Physics, Chinese Academy of Sciences, 100049 Beijing, China}
\altaffiltext{8}{University of Chinese Academy of Sciences, Chinese Academy of Sciences, 100049 Beijing, China}

%author[0000-0003-3188-9079]{Lingda Kong}
%\affil{Institut f{\"u}r Astronomie und Astrophysik, Kepler Center for Astro and Particle Physics, Eberhard Karls, Universit{\"a}t, Sand 1, D-72076 T{\"u}bingen, Germany}

\begin{abstract}
We present an analysis of the relativistic reflection spectra of GX~339--4 during the hard-to-soft transition of its 2021 outburst observed by \textit{Insight}--HXMT. The strong relativistic reflection signatures in the data suggest a high black hole spin ($a_*>0.86$) and an intermediate disk inclination angle ($i \approx$ 35$^\circ$--43$^\circ$) of the system. The transition is accompanied by an increasing temperature of the disk and a softening of the corona emission while the inner disk radius remains stable. Assuming a lamppost geometry, the corona height is also found to stay close to the black hole across the state transition. If we include the Comptonization of the reflection spectrum, the scattering fraction parameter is found to decrease during the state transition. We also perform an analysis with a reflection model designed for hot accretion disks of stellar mass black holes where the surface of the innermost accretion disk is illuminated by emission from the corona and the thermal disk below. Our results support the scenario in which the state transition is associated with variations in the corona properties.
%A super-solar iron abundance is always required to fit the reflection spectra.
\end{abstract}


%%%%%%%%%%%%%%%%%%%%%%%%%%%%%%%

\section{Introduction}

%%%%%%%%%%%%%%%%%%%%%%%%%%%%%%%

% The outburst of black hole XRB

The outbursts of black hole X-ray binaries (XRBs) are characterized by orders of magnitude change in the X-ray luminosity \citep{Remillard2006}. It is generally believed that there are dramatic changes in the geometry of the accretion flow during the outburst \citep[e.g.][]{Zdziarski2004, Done2007, Plant2015}. If plotting on the hardness intensity diagram (HID), the outburst often moves anticlockwise to form a q-shaped pattern \citep[e.g.][]{Fender2004, Belloni2005}. At the early stage of the outburst, the source luminosity is low and its X-ray spectrum is usually dominated by a hard power-law component, which is thought to be originated from inverse Compton scattering of seed photons by a hot ($\sim$ 100 keV) corona \citep[e.g.][]{Dove1997, Zdziarski2004}. This is the so called low hard (LH) state that is located at the lower right corner of the HID. At this phase, the source spectrum remains hard as it gets brighter. Moreover, a high-energy cut-off of the broadband X-ray spectrum is commonly observed and is thought to be determined by the coronal temperature. The energy of this cut-off is found to decease with the source luminosity \citep[e.g.][but see \citealt{Yan2020} for a opposite correlation at even lower luminosities]{Yamaoka2005,Miyakawa2008,Joinet2008}.

% The transition between hard and soft states: time scale, spectral changes, jet, truncation disk, moving inward, hot inflow

At some point, the source makes a fast transition from the hard state to a thermal-emission-dominated soft state through the intermediate states \citep[e.g.][]{Homan2005, Belloni2010}. The transition, which can happen in a wide range of luminosities, often completes in a few days and is associated with softening of the source spectrum. This softening is a result of the increasing contribution from the disk thermal emission and steepening of the corona emission. Overall, the X-ray spectrum appears as pivoting around certain energy \citep[e.g.][]{Del-Santo2008}. The cut-off energy tends to rise during the hard to soft transition \citep[e.g.][]{Motta2009MNRAS.400.1603M} and in some cases, no high-energy cut-off is found up to $\sim$ 1~MeV in the soft state \citep[e.g.][]{Tomsick1999,Belloni2006}.

%but the evolution of this parameter is complicate. Both increasing and decreasing corona temperature during the state transition have been reported \citep[e.g.][]{Del-Santo2008,Motta2009MNRAS.400.1603M}

A commonly accepted scenario to explain the state transition is the truncated disk model \citep[e.g.][]{Esin1997}. In this scenario, the standard geometrically thin and optically thick accretion disk \citep{Shakura1973} is truncated at large radius in the low hard state when the accretion rate is low. A hot radiatively inefficient accretion flow is filled between the truncation radius and the black hole \citep[e.g.][]{Narayan2005}. As the accretion rate increases, the disk moves closer to the black hole, which produces stronger disk thermal emission. X-ray observations of XRBs have provided good evidence that the disk is truncated at large radius in low luminosities \citep[e.g.][]{Tomsick2009} and reaches the innermost stable circular orbit (ISCO) in the high flux soft state \citep[e.g.][]{Gierlinski2004, Steiner2010}. However, it remains to be known at which point the disk exactly reaches the ISCO and what mechanism triggers the state transition. Moreover, in some cases, the measurements of the size of inner edge are found to be controversial and model dependent \citep[e.g.][]{Dzielak2019, Mahmoud2019, Zdziarski2021ApJ...906...69Z}.

%The soft-to-hard transition happens at a lower luminosity than the hard-to-soft transition, which produces the hysteresis effect.

% relativistic reflection as a probe of the standard accretion disk, assumptions, simplifications

Relativistic reflection component, namely the reflected corona emission by the optically thick accretion disk extending close to the black hole, is a powerful tool to probe the spacetime properties \citep[e.g.][]{Reynolds2014, Bambi2017, Liu2019, Bambi2021} and the geometry of the accreting system \citep[e.g.][]{Garcia2015, Wilkins2015, Wang2020, Szanecki2020}. One of the main features of the relativistic reflection component is the broad iron line around 6.4 keV. The broad line profile is a result of skewed iron K lines due to the strong relativistic effect near the black hole \citep[e.g. Doppler effect, light bending and gravitational redshift,][]{Fabian2000, Dauser2010}. This is why this broad line feature can be used to map the innermost regions around the accreting black hole \citep[e.g.][]{Fabian1989}.

% GX 339-4, general information, the 2021 outburst

\src{} is a classical low mass X-ray binary located at 8--12~kpc with a black hole mass of 4--11~$M_{\odot}$ \citep{Zdziarski2019}. Strong relativistic reflection signatures have been found in this source in the hard and soft states \citep[e.g.][]{Garcia2015, Miller2004, Liu2022}. Previous studies have found that the black hole in \src{} has a very high spin \citep[$a_*\sim0.95$,][]{Garcia2015, Parker2016}. The inclination angle of the accretion disk should have an intermediate value \citep{Furst2015, Parker2016}. The source goes into bright outburst every a few years and thus is an ideal object for the study of the accretion flow through different spectra states. Recently, the hard-to-soft transition of \src{} has been studied by \cite{Sridhar2020} with \textit{RXTE}/PCA data based on reflection analysis. The inner edge of the accretion disk was found to be close to ISCO during the transition. Starting from February 2021, \src{} went into a new outburst that lasted for a few months. The hard to soft transition of this outburst has been captured by the Chinese X-ray satellite Hard X-ray Modulation Telescope \citep[dubbed \hxmt{},][]{Zhang2014}. Therefore, we study the \hxmt{} data to understand the evolution of the accretion geometry.

% We note that \hxmt{} has a better energy resolution (150 eV at 6 keV) and covers a broader energy band than \textit{RXTE}.

%It would be useful to see how \hxmt{} data would improve our understanding of the accretion geometry.

In this paper, we aim at studying the relativistic reflection spectra of \src{} during its 2021 outburst with \hxmt{} data. In Sec.~\ref{obs}, we describe the data reduction process. The spectral fitting results are presented in Sec.~\ref{s-ana}. We discuss the results in Sec.~\ref{s-dis}.

\section{Observation and data reduction}\label{obs}

% Data selection

\begin{figure*}
    \centering
    \includegraphics[width=0.49\linewidth]{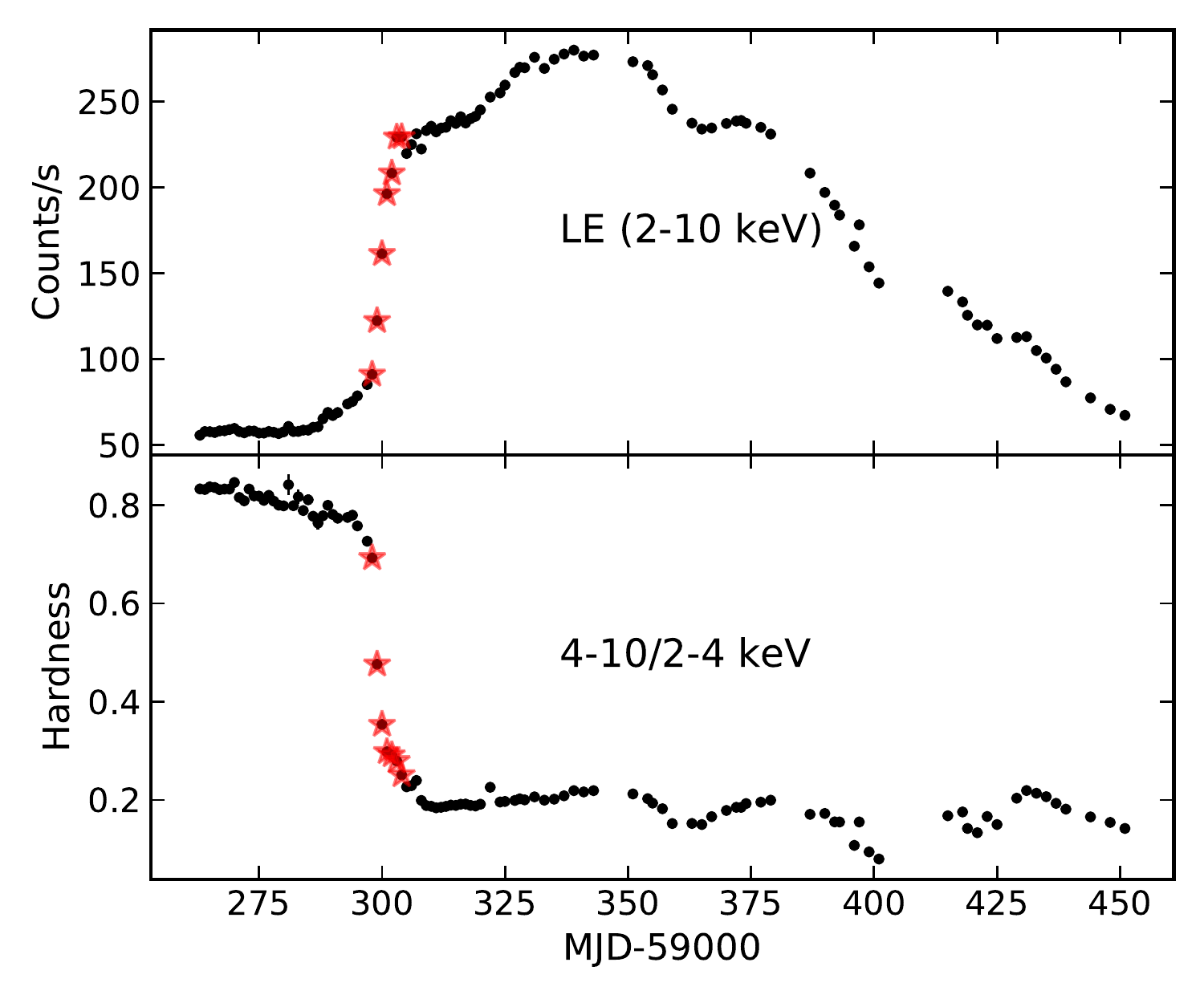}
    \includegraphics[width=0.49\linewidth]{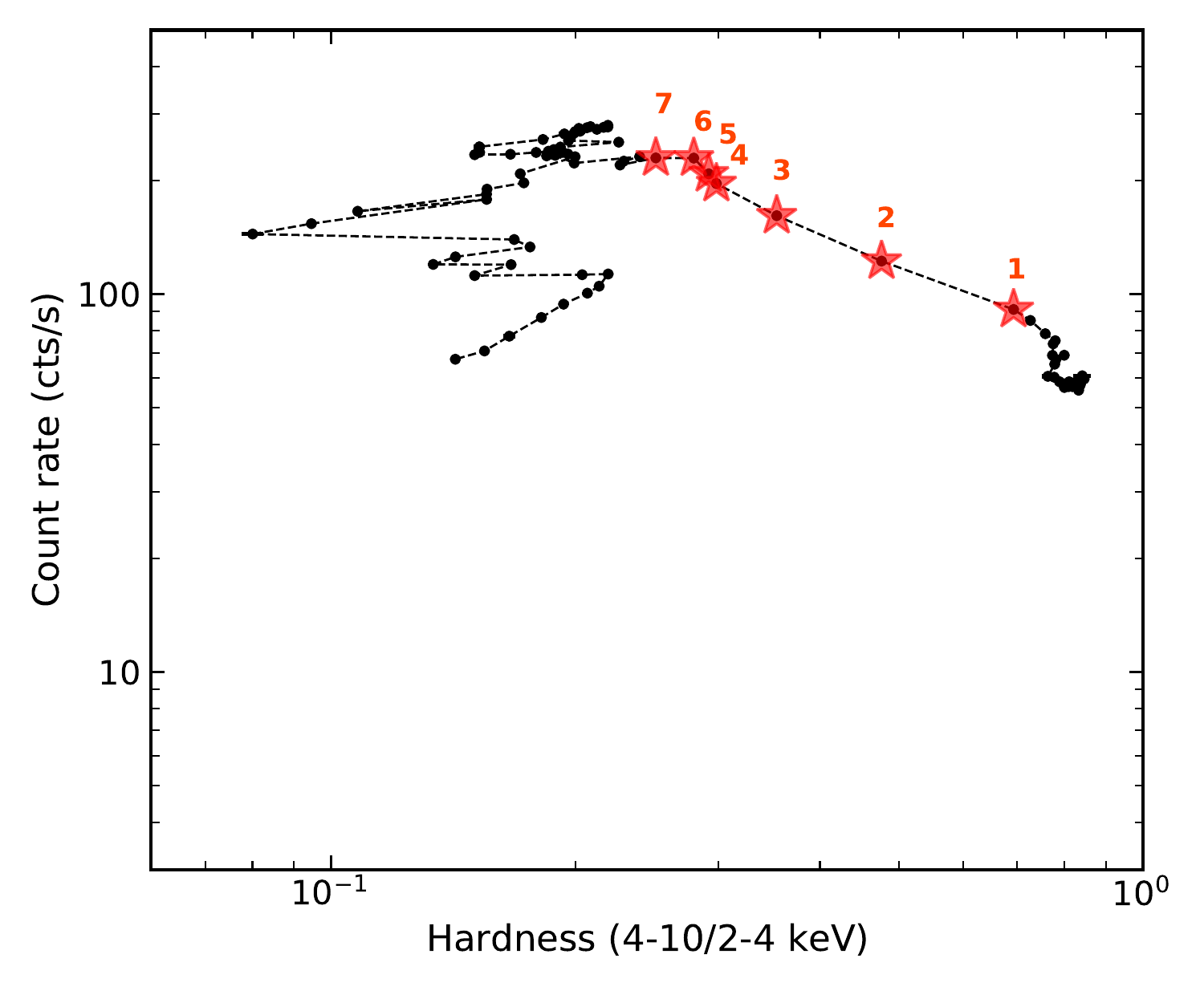}
    \caption{(left): The observation log of GX~339--4 from \hxmt{} Low Energy (LE) instruments. The hardness is defined as the ratio between count rates in 4--10 keV and 2--4 keV. (right): The hardness intensity diagram of GX~339--4 from \hxmt{}. Each point represents data from one day and is marked with the epoch number. The red color denotes the observations analyzed in this work.}
    \label{log}
\end{figure*}

\hxmt{} monitored the 2021 outburst of \src{} with observations on nearly every day (see Fig.\ref{log}). Starting from 2021 March 25, the source made a fast transition from the hard state to the soft state in about one week. The soft to hard transition, which should happen at a lower luminosity, was not captured because of the high background of \hxmt{}. The purpose of this work is to study the evolution of the system during the state transition. Therefore, we select \hxmt{} observation data during the transition and investigate their broadband energy spectra. Details about the selected \hxmt{} observations are listed in the appendix.

% Data reduction

The low-energy (LE), medium-energy (ME) and high-energy (HE) detectors onboard \hxmt{} cover the energy range of 1-250 keV \citep{Chen2020, Cao2020, Liu2020, Zhang2020}. The energy spectra are extracted following the HXMT Data Reduction Guide v2.04\footnote{The guide can be found from here: \url{http://hxmtweb.ihep.ac.cn/SoftDoc/496.jhtml}. We follow the extraction approach provided here: \url{https://code.ihep.ac.cn/jldirac/insight-hxmt-code-collection/-/tree/master/version2.04}.} and using the software \texttt{HXMTDAS} v2.04. The background spectra are estimated using the scripts \texttt{hebkgmap}, \texttt{mebkgmap} and \texttt{lebkgmap} \citep{Liao2020a, Guo2020, Liao2020b}. As indicated in Tab.~\ref{info-obs}, short exposures on the same day are merged to increase the signal to noise ratio. Before merging the data, we first check the lightcurve and hardness ratio and confirm that there are no significant flares or dips. In the end, we obtain energy spectra for seven epochs. The spectra are then binned to ensure a minimal signal to noise ratio of 25 before fitting them to physical models. We use data in 2--10 keV, 10--25 keV and 30-100 keV bands for LE, ME and HE respectively.

%%%%%%%%%%%%%%%%%%%%%%%%%%%%%%%

\section{Spectral analysis}\label{s-ana}

% Softwares, abundances
Spectral fittings are conducted with \texttt{XSPEC} v12.11.1 \citep{xspec}. We implement element abundances of \cite{Wilms2000} and cross-sections of \cite{Verner1996}. $\chi^2$ statistics are used to find the best-fit values and uncertainties (at 90\% confidence level unless otherwise specified) of the model parameters.

\subsection{Relativistic reflection features}

As the first step, we fit the seven spectra simultaneously with a simple absorbed continuum model which consists of a multicolor disk component \citep[\texttt{diskbb;}][]{Mitsuda1984}, a power-law component with a high energy cutoff (\texttt{cutoffpl}) and a Galactic absorption model \citep[\texttt{tbabs;}][]{Wilms2000}. In \texttt{XSPEC} notation, the model can be written as \texttt{constant*tbabs*(diskbb + cutoffpl)}. The \texttt{constant} is required to quantify the cross-normalization between the three instruments onboard \hxmt{}. The column density ($N_{\rm H}$) of the \texttt{tbabs} is set as a free parameter and tied across all spectra. 

\begin{figure*}
    \centering
    \includegraphics[width=\linewidth]{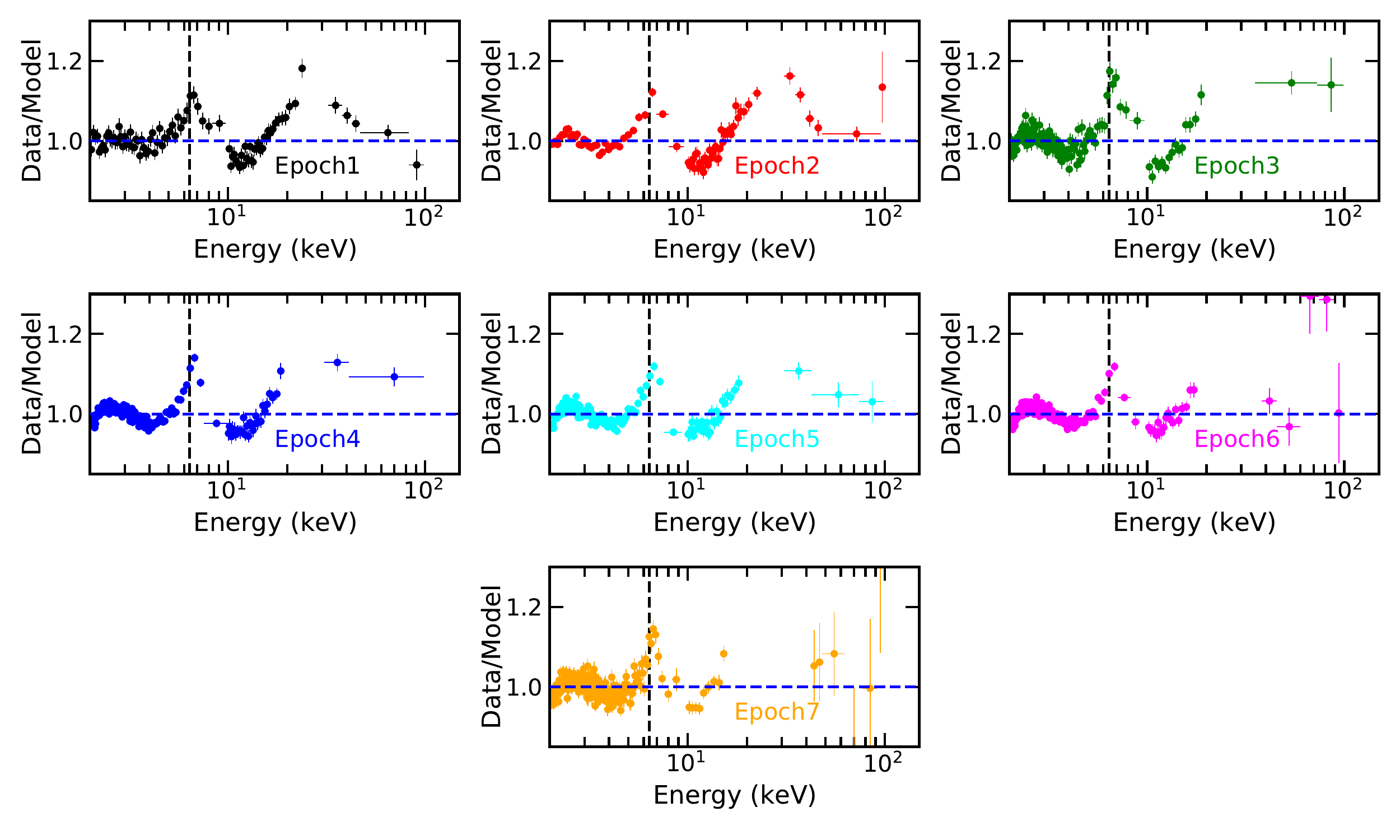}
    \caption{Data to model ratios for an absorbed continuum model \texttt{tbabs*(diskbb+cutoffpl)} for the seven epochs. The vertical line marks the position of 6.4 keV. Data are rebinned for visual clarity.}
    \label{ironline}
\end{figure*}

The ratios between the data and the best-fit models for the seven spectra are shown in Fig.~\ref{ironline}, from which we can identify the missing model components. We can see that the common features in Fig.~\ref{ironline} are a broad excess around 6--7 keV and a hump around 30 keV. These features are commonly seen in the X-ray spectra of black hole XRBs \citep[e.g.][]{Miller2013, Garcia2015, Jiang2019b} and can be explained by the corona emission being reflected by the optically thick accretion disk \citep[e.g.][]{Ross1999}. The position of the neutral iron K$\alpha$ fluorescent line (6.4 keV) is marked with vertical dashed line in Fig.~\ref{ironline}. The broad emission profile around this line is a result of relativistically skewed iron K lines. Moreover, the electron down scattering of high energy photons and photoelectric absorption on the disk leads to the ``Compton hump" around 30 keV \citep{George1991}.

\subsection{Spectral fitting with reflection models}

To model the relativistic reflection features, we implement the widely used reflection model \texttt{relxill} v1.4.3 \citep{Garcia2014}. The full model reads as \texttt{constant*tbabs*(diskbb+relxill+cutoffpl)} (Model 1). With this model, we are able to measure parameters of the system like the black hole spin ($a_*$), the inclination angle ($i$) of the accretion disk with respect to our line of sight and the size of the disk inner edge ($R_{\rm in}$). The seven spectra are fitted simultaneously. The column density of the Galactic absorption ($N_{\rm H}$), the black hole spin, the inclination angle and the iron abundance of the system should not change on the short timescale we are seeing. Therefore, these parameters are linked across all spectra during the spectral fitting. The reflection fraction parameter of \texttt{relxill} is set to $-1$ so that the model returns only the reflection component. The outer radius of the disk $R_{\rm out}$ is fixed at 400 $R_{\rm g}$ (where $R_{\rm g}=GM/c^2$ is the gravitational radius, $M$ is the black hole mass).

\subsubsection{Lamppost geometry}

The emissivity profile, defined as the radial dependence of the intensity of reflected emission, is a vital component of the reflection model. In principle, if the corona geometry is known, the emissivity profile can be self-consistently calculated. Therefore, we implement the \texttt{relxilllp} (Model 1A), which is one of the models in the \texttt{relxill} package. The model assumes a lamppost geometry \citep{Dauser2013}, in which the corona is a point source above the black hole at certain height ($h$). This model offers a good fit with $\chi^2$/d.o.f=3697.6/3384. The best-fit values are shown in Tab.~\ref{lp}. The spectral components and residuals of the best-fit model are shown in Fig~\ref{resi}, from which we do not see significant unresolved features. With this model, we can fit with both the spin parameter and the inner disk radius free. There is usually strong degeneracy between the two parameters \citep[e.g.][]{Dauser2013}. Our data can break this degeneracy because the disk extends very close to the ISCO \citep{Fabian2014}. If we fix the spin parameter at the maximum value the model allows ($a_*=0.998$), $R_{\rm in}$ is still not constrained for Epoch~1 but is smaller than twice of the ISCO from Epoch~4 to Epoch~7 ($\chi^2$/d.o.f=3697.6/3385). Moreover, if we set $R_{\rm in}$ at ISCO for all epochs, the spin is measured to be $0.93\pm0.01$ ($\chi^2$/d.o.f=3704.0/3391).

\begin{table*}
    \centering
    \caption{Best-fit values with the lamppost geometry (Model 1A).}
    \label{lp}
    \renewcommand\arraystretch{2.0}
    %\resizebox{0.45\textwidth}{4.5cm}{
    \begin{tabular}{lcccccccc}
        \hline\hline
        Component       & Parameter & Epoch 1 & Epoch 2 & Epoch 3 & Epoch 4 & Epoch 5 & Epoch 6 & Epoch 7 \\
        \hline
        \texttt{tbabs} & $N_{\rm H}$ (10$^{22}$ cm$^{-2}$) & \multicolumn{7}{c}{$0.99_{-0.06}^{+0.04}$} \\
        \hline
        \texttt{diskbb} & $T_{\rm in}$ (keV) & $0.50_{-0.03}^{+0.03}$ & $0.548_{-0.01}^{+0.014}$ & $0.656_{-0.008}^{+0.008}$ & $0.725_{-0.008}^{+0.011}$ & $0.725_{-0.009}^{+0.01}$ & $0.766_{-0.007}^{+0.006}$ & $0.757_{-0.007}^{+0.011}$\\
        \hline
        \texttt{relxilllp} & $a_*$ & \multicolumn{7}{c}{$>0.95$} \\
        & $h$ ($R_{\rm g}$) & $3.5_{-P}^{+P}$ & $3.1_{-0.5}^{+0.5}$ & $2.0_{-P}^{+4}$ & $2.20_{-0.1}^{+0.21}$ & $2.06_{-P}^{+0.18}$ & $2.35_{-0.29}^{+0.28}$ & $2.03_{-P}^{+0.24}$\\
        & Incl & \multicolumn{7}{c}{$42.0_{-1.4}^{+1.3}$} \\
        & $A_{\rm Fe}$ (solar) & \multicolumn{7}{c}{$10.0_{-0.6}^{+P}$} \\
        & $R_{\rm in}$ (ISCO) & $70_{-P}^{+P}$ & $1.4_{-P}^{+0.7}$ & $3.2_{-1.9}^{+1.2}$ & $1.56_{-0.07}^{+0.08}$ & $1.44_{-0.09}^{+0.08}$ & $1.62_{-0.13}^{+0.12}$ & $1.6_{-0.1}^{+0.1}$\\
        & $\log(\xi)$ & $4.29_{-0.06}^{+0.06}$ & $3.89_{-0.23}^{+0.17}$ & $4.51_{-0.11}^{+0.06}$ & $3.71_{-0.08}^{+0.09}$ & $4.06_{-0.2}^{+0.22}$ & $4.14_{-0.19}^{+0.2}$ & $3.79_{-0.12}^{+0.13}$\\
        \hline
        \texttt{cutoffpl} & $\Gamma$ & $1.696_{-0.013}^{+0.018}$ & $2.087_{-0.019}^{+0.028}$ & $2.260_{-0.019}^{+0.06}$ & $2.460_{-0.03}^{+0.021}$ & $2.451_{-0.019}^{+0.024}$ & $2.511_{-0.024}^{+0.028}$ & $2.55_{-0.08}^{+0.1}$\\
        & $E_{\rm cut}$ (keV) & $150_{-20}^{+20}$ & $160_{-20}^{+40}$ & $>400$ & $>400$ & $>330$ & $>410$ & $130_{-60}^{+50}$\\
        \hline
        \texttt{cflux} & (10$^{-8}$~erg~cm$^{-2}$~s$^{-1}$)\\
        $F_{\rm diskbb}$ & & $0.23_{-0.03}^{+0.03}$ & $0.552_{-0.028}^{+0.019}$ & $0.935_{-0.023}^{+0.019}$ & $1.314_{-0.017}^{+0.03}$ & $1.355_{-0.025}^{+0.021}$ & $1.603_{-0.022}^{+0.019}$ & $1.85_{-0.05}^{+0.03}$\\ 
        $F_{\rm relxilllp}$ & & $0.77_{-0.06}^{+0.10}$ & $0.74_{-0.08}^{+0.08}$ & $1.43_{-0.11}^{+0.13}$ & $1.52_{-0.12}^{+0.18}$ & $1.52_{-0.12}^{+0.2}$ & $1.64_{-0.14}^{+0.19}$ & $1.9_{-0.5}^{+1.1}$\\
        $F_{\rm cutoffpl}$ & & $0.71_{-0.07}^{+0.05}$ & $1.28_{-0.15}^{+0.13}$ & $0.94_{-0.20}^{+0.10}$ & $2.04_{-0.21}^{+0.13}$ & $1.77_{-0.12}^{+0.1}$ & $1.90_{-0.27}^{+0.09}$ & $1.6_{-0.3}^{+0.4}$\\
        \hline
        & $L_{\rm 0.1-100 keV}/L_{\rm Edd}$ (\%) & 10.9 & 16.4 & 21.1 & 31.1 & 29.6 & 32.8 & 34.1 \\
        \hline
        \texttt{constant} & ME/LE & \multicolumn{7}{c}{$1.023_{-0.014}^{+0.013}$} \\
        & HE/LE & \multicolumn{7}{c}{$1.110_{-0.023}^{+0.019}$} \\ 
        \hline
                        & $\chi^2$/d.o.f &  \multicolumn{7}{c}{3697.6/3384} \\
        \hline\hline
    \end{tabular}
    %}
    \textit{Note.} Best-fit parameters for the model \texttt{tbabs*(diskbb+relxilllp+cutoffpl)}. Parameters with $^*$ are fixed during the fit and the symbol $P$ denotes the upper or lower boundary. Parameters that are only shown for Epoch~4 are tied across all observations. The flux of each spectral components are calculated in the 0.1-100 keV band using the \texttt{cflux} model. The absorption corrected X-ray flux ($L_{\rm 0.1-100 keV}$) is calculated with the \texttt{flux} command in \texttt{XSPEC}.
\end{table*}

\subsubsection{Broken power-law emissivity}

The other method commonly used to deal with the emissivity profile is to fit it phenomenologically with a broken power-law (i.e., $\epsilon\propto 1/r^{q_{\rm in}}$ for $R_{\rm in}<r<R_{\rm br}$ and $\epsilon\propto 1/r^{q_{\rm out}}$ for $R_{\rm br}<r<R_{\rm out}$ where $R_{\rm br}$ is the breaking radius). We test both the power-law emissivity ($q_{\rm in}=q_{\rm out}$) and the broken power-law emissivity (implemented with the model \texttt{relxill}) to look for a better description of the data. In the latter case, we fix the outer index fixed at 3 ($q_{\rm out}=3$), which is the value predicted by Newtonian limit for a compact corona. Moreover, for the first three epochs, the inner index ($q_{\rm in}$) and the breaking radius ($R_{\rm br}$) can not be constrained by the data. Therefore, we fix $q_{\rm in}$ at 3 for those data. The broken power-law emissivity provides a better statistics ($\chi^2/\nu=3673.1/3383$), with the $\chi^2$ lowered by 13 with 2 more free parameters. The best-fit values for the broken power-law emissivity (Model 1B) are shown in Tab.~\ref{bpo}. 

% We note that releasing $q_{\rm out}$ only improves the $\chi^2$ by 13 and there is no significant impact on other parameters.

\subsubsection{High-density reflection}

The disk density ($n_{\rm e}$) of the models above is fixed at 10$^{15}$~cm$^{-3}$. This value might be appropriate for massive black holes but XRBs with a stellar mass black hole are known to have a higher density \citep[e.g.][]{Jiang2019b,Jiang2019}. Therefore, we implement the \texttt{relxillD} model (assuming a broken power-law emissivity, Model 1C) in which $\log(n_{\rm e})$ is allowed to vary between 15 and 19 \citep{Garcia2016}. The best-fit parameters are shown in Tab.~\ref{relxilld}. Compared to the model above, the high-density model slightly improves the $\chi^2$ by 9 with 7 more degrees of freedom and provides consistent constraints on the spin parameter and the inclination angle. Five out of the seven observations require a higher density than 10$^{15}$~cm$^{-3}$. The measurements on the ionization parameter are systematically lowered. This is expected since a higher density leads to stronger soft X-ray emission, which is similar to the effect of a higher ionization parameter.

\subsubsection{The Comptonized continuum}

The models tested above provide good fits to the data. However, the column density of the \texttt{tbabs} component is higher than what reported in previous spectral analysis \citep[e.g. $7.7\pm 0.2\times 10^{21}$ cm$^{-2}$ in][with the same versions of abundances and cross-sections]{Parker2016}. This is recovered if we replace \texttt{cutoffpl} with a physically motivated Comptonization model \texttt{nthcomp} \citep{Zdziarski1996, Zycki1999}. The reflection component is also changed to the corresponding flavor \texttt{relxillcp} (Model 2). We link the seed photon temperature for the \texttt{nthcomp} model to the $T_{\rm in}$ parameter in \texttt{diskbb}. The best-fit parameters for this model are shown in Tab.~\ref{relxillcp}. Compared with \texttt{relxill}, this model increases the statistics  by $\Delta\chi^2=64$ but lowers the column density of the Galactic absorption to $7.3\pm 0.6\times 10^{21}$ cm$^{-2}$. This is because the Comptonization model considers cutoffs at both low and high energy ends. Therefore, the X-ray luminosity estimated from the Comptonization model should be more realistic than that from the model with \texttt{cutoffpl}. We also note that our data cannot well constrain the coronal temperature (or the high energy cutoff).

\subsubsection{Distant reflector}

The spectra of \src{} may require a distant reflection component to fit a possible narrow-line feature \citep[e.g.][]{Wang2018, Sridhar2020}. To test this possibility, we add a \texttt{xillver} \citep{Garcia2010, Garcia2013} component to Model 1A. The relativistic reflection model assumes a broken power-law emissivity profile. The ionization parameter of this distance reflection component is fixed at $\log(\xi)=0$. The other parameters of the distance reflector are linked to the corresponding values in the relativistic reflection component. Adding this component improves the $\chi^2$ by only 10 with 7 more free parameters. 

To test if our data quality is good enough to distinguish such a distant reflector, we simulate 1000 spectra with a model that includes a distant reflection component. The response files, exposure time and model parameters (Tab.~\ref{lp}) are taken from our Epoch 5 which has the longest exposure. The flux (0.1--100~keV) of the distant reflection component is set to be 2\% \citep[e.g.][]{Sridhar2020} of the relativistic component. Then we fit the spectra with two models: one without the distant reflection and the other includes it. The latter model always gives a better statistics with $\Delta\chi^2>50$ and one more degrees of freedom, which indicates that the distant component can be detected if it exists. Therefore, we conclude that, for the observation with the longest exposure time, the distant reflection component can be detected if its flux is larger than 2\% of the relativistic component. The reason why we do not detect the distant component in our analysis could be that its flux is lower than the limit provided by our data quality. Similarly, in \cite{Jiang2019b}, the intermediate state of \src{} requires no distance reflection.

\begin{figure*}[htbp]
    \centering
    \includegraphics[width=0.33\linewidth]{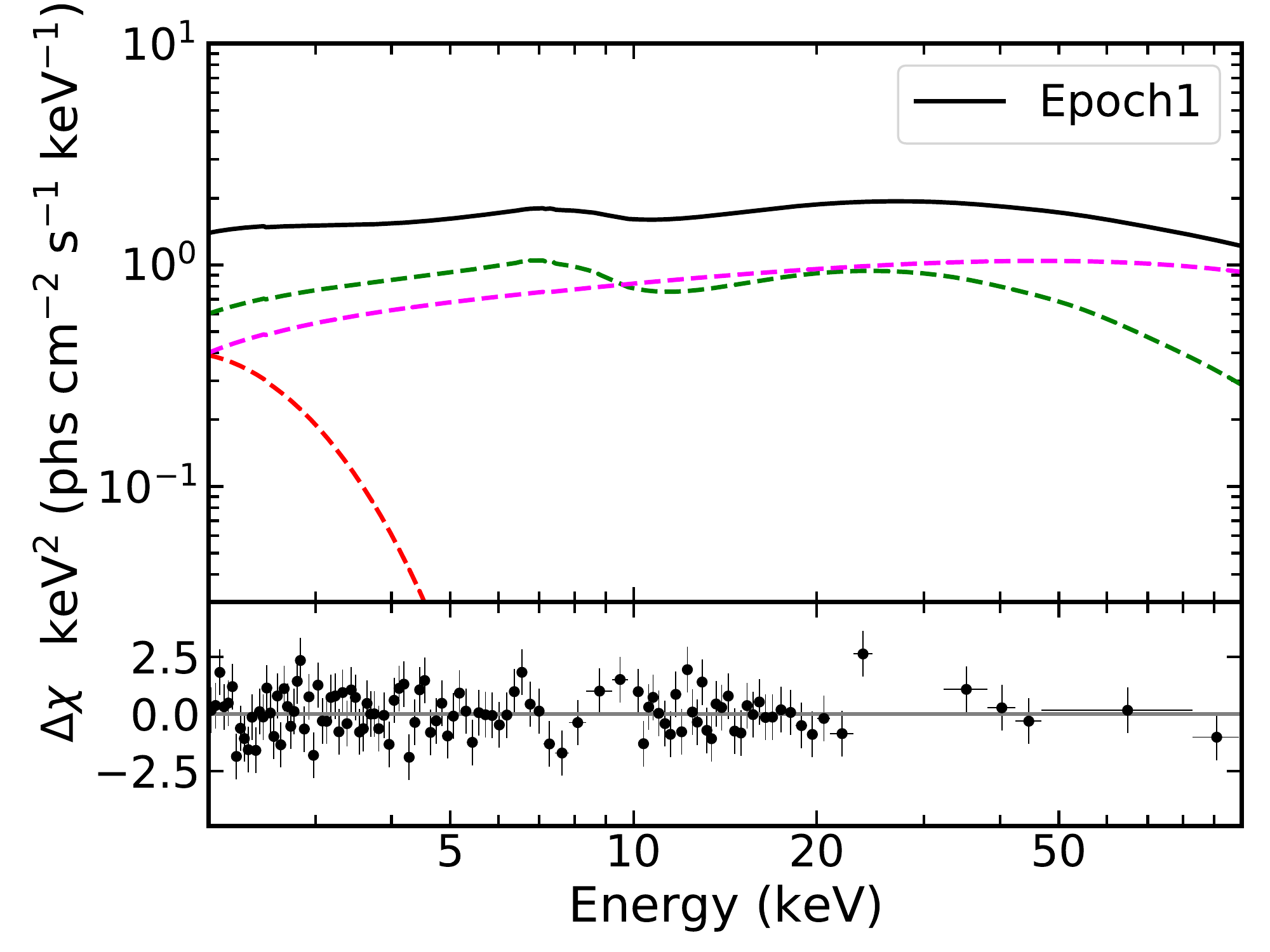}
    \includegraphics[width=0.33\linewidth]{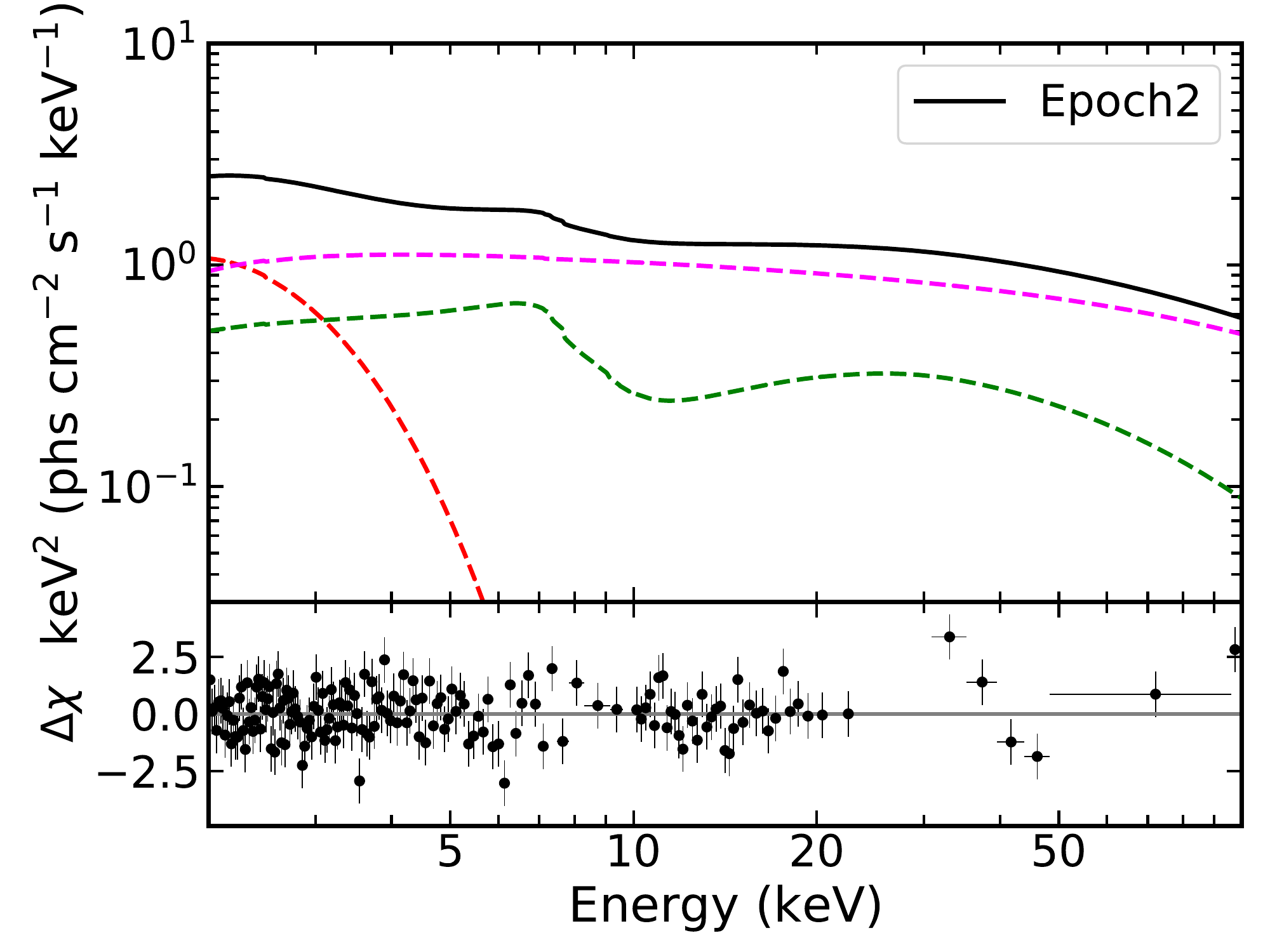}
    \includegraphics[width=0.33\linewidth]{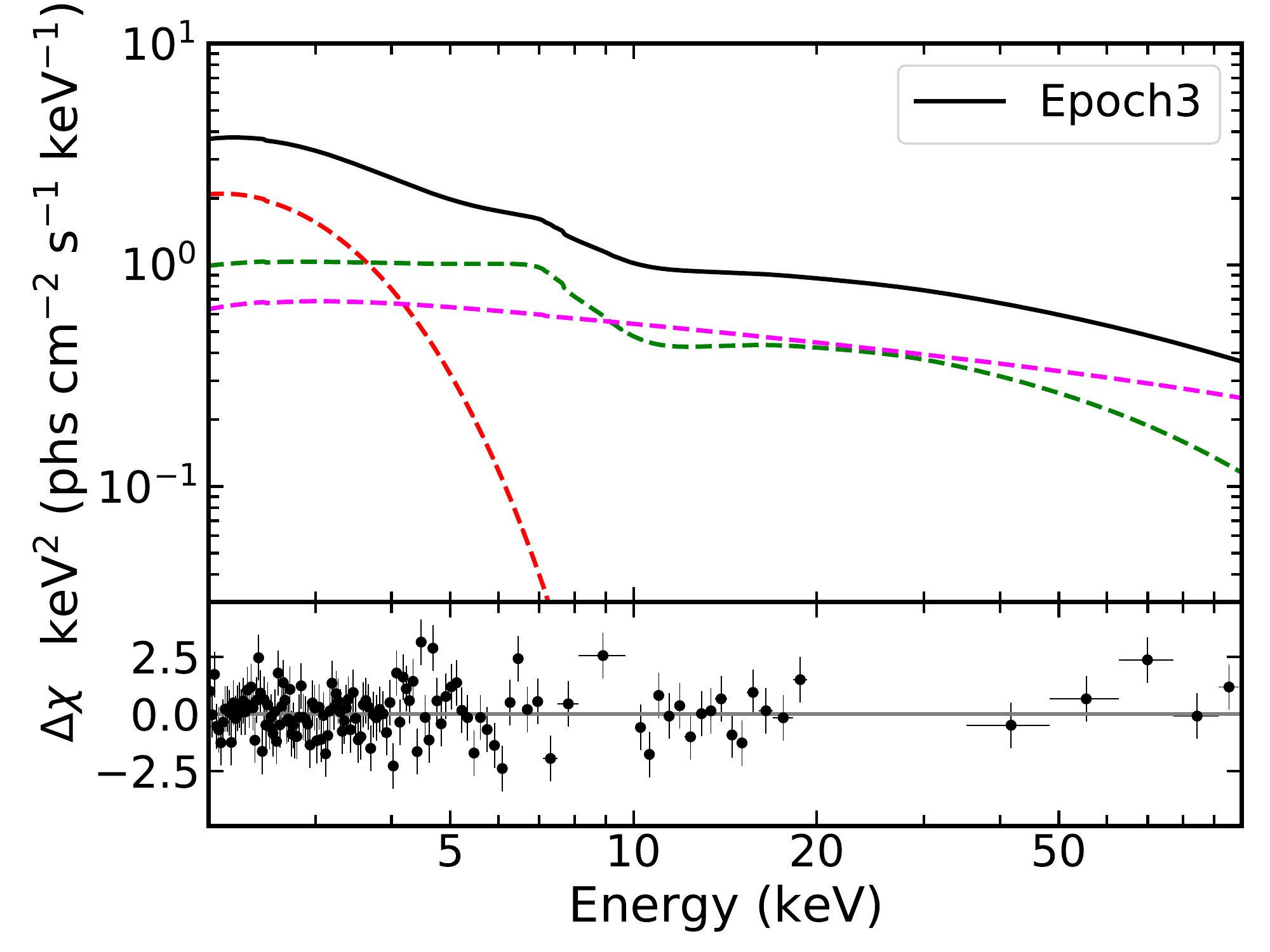}\\
    \includegraphics[width=0.33\linewidth]{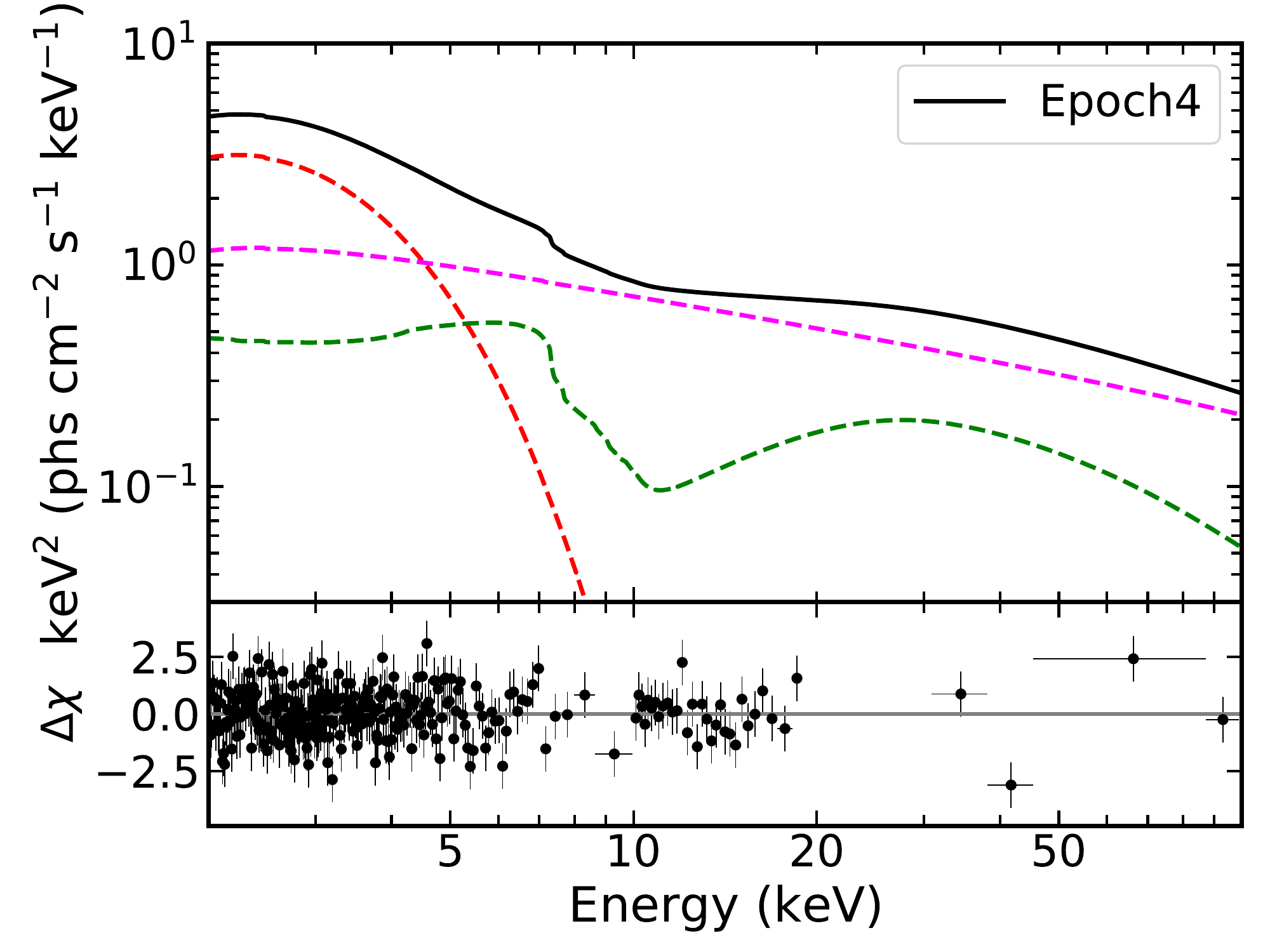}
    \includegraphics[width=0.33\linewidth]{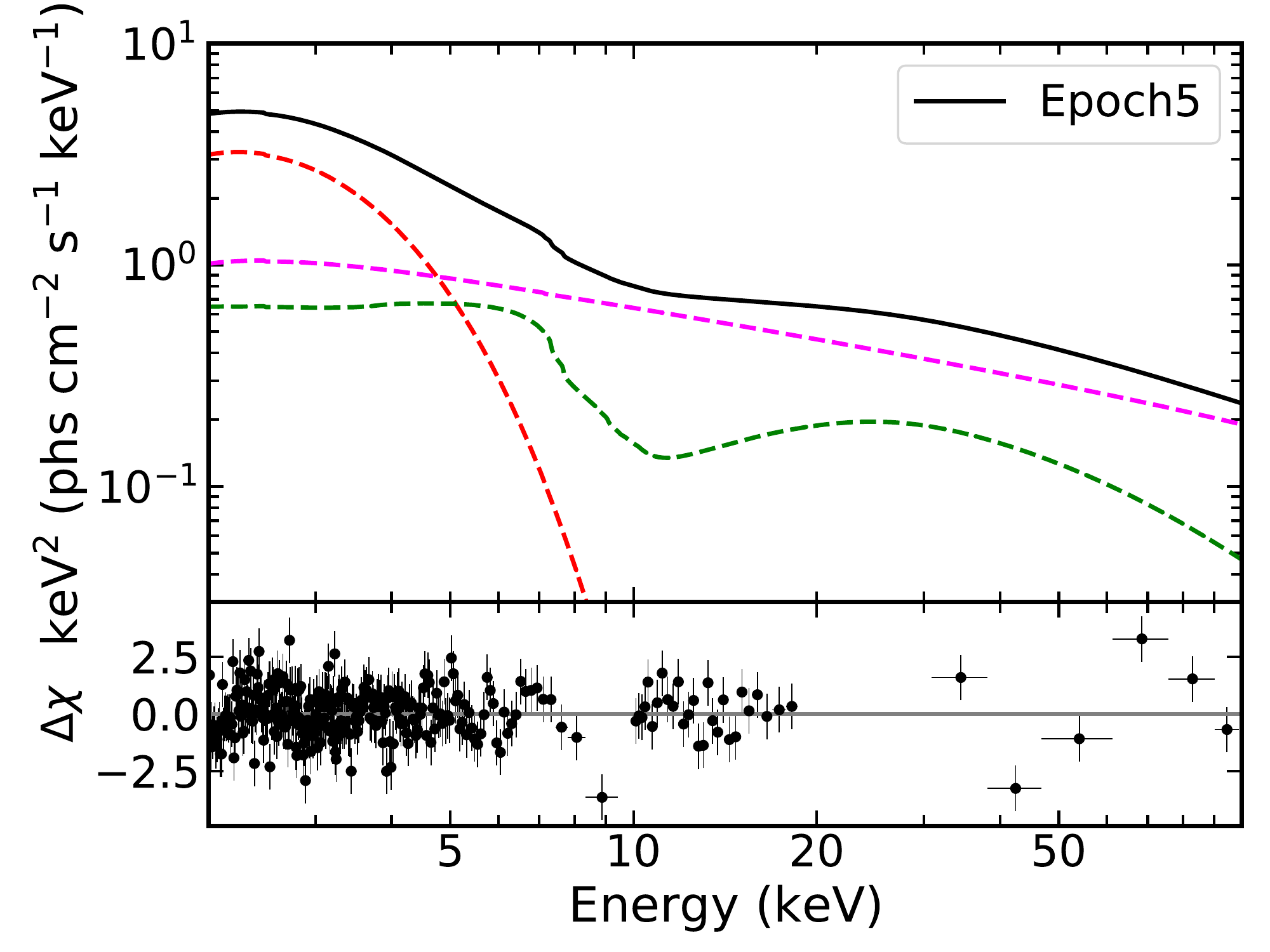}
    \includegraphics[width=0.33\linewidth]{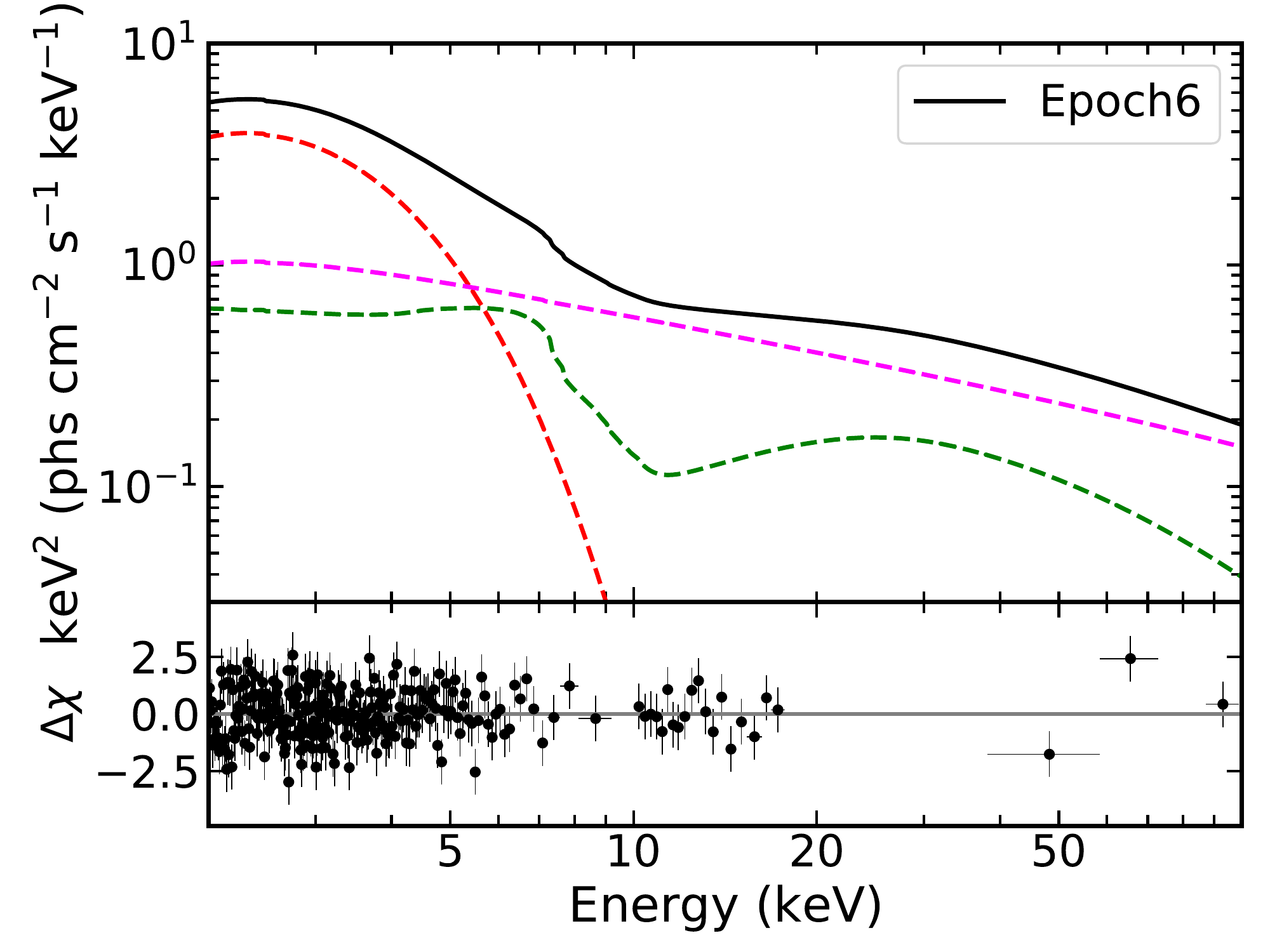}\\
    \includegraphics[width=0.33\linewidth]{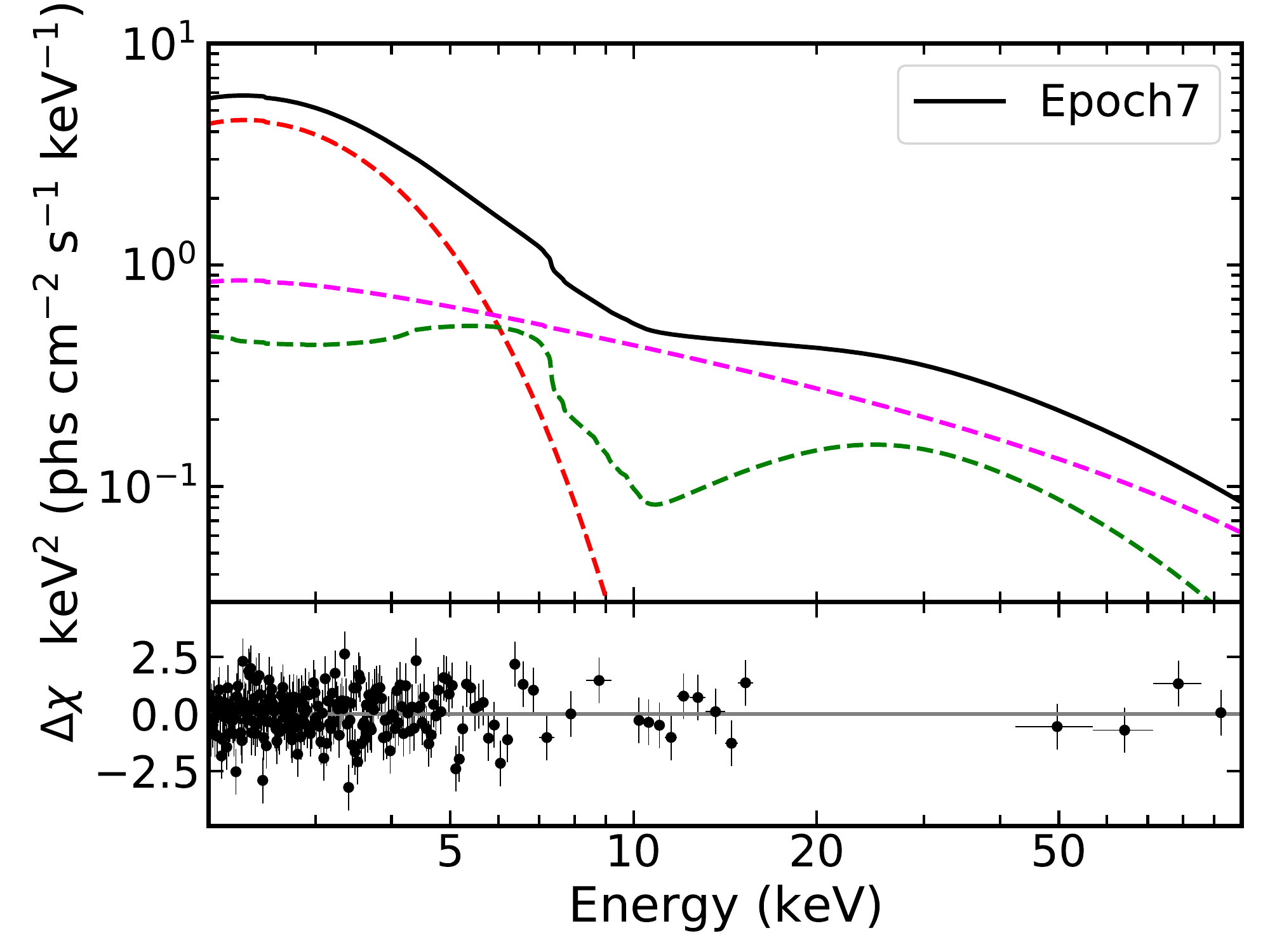}
    \caption{The best-fit model components for the seven epochs and the corresponding residuals with the model \texttt{tbabs*(diskbb+relxilllp+cutoffpl)}. The black solid line represents the total model. The magenta, red and green dashed lines are for the corona emission, the disk component and the reflected component respectively.}
    \label{resi}
\end{figure*}

\subsubsection{Comptonization of the reflection component}
\label{comp}

The power-law emission is the result of inverse Compton scattering of the disk photons by the hot electrons \citep[but see][]{Sridhar2021MNRAS.507.5625S, Sridhar2022MNRAS} in the corona. Some of the reflected emission should also intercept the corona and be Comptonized. Therefore, a self-consistent model should take the Comptonized reflection spectrum into account \citep{Steiner2017, Wang2018}. We consider this scenario with the Compton scattering kernel \texttt{simplcut} \citep{Steiner2017}. The full model reads as \texttt{tbabs*simplcut*(diskbb+relxillcp)} (Model 3). The photon index ($\Gamma$) and electron temperature ($kT_{\rm e}$) of \texttt{simplcut} are linked to the corresponding parameters in \texttt{relxillcp}. There are two more parameters in the Compton scattering kernel: the reflection fraction $R_f$ and the scattered fraction $f_{\rm sc}$. We fix the reflection fraction parameter at the flux ratio between the reflection and power-law components in the 20--40 keV band derived from Model 2 and let the scattered fraction parameter free to vary.

With this model, we are assuming an equal fraction of the disk photons and the reflected photons are scattered by the corona. The best-fit parameters are shown in Tab.~\ref{simplcut}. We note that including the effect of Comptonization of the disk and reflection photons does not change significantly the measurements of parameters like $T_{\rm in}$, $\Gamma$ and $R_{\rm in}$. The measurements of the spin and inclination parameters are also not affected. The scattered fraction decreases from 0.12 to 0.07, which may indicate changes of the disk-corona system during the transition.

\subsubsection{The \texttt{refhiden} model}

In the \texttt{relxill} model, the ionization balance on the disk is calculated using only the incident corona emission. In reality, we would expect that the thermal emission from the disk can also affect the ionization state. This is particularly important in XRBs since the disk temperature of XRBs can be very high ($\sim$ 1~keV) and the thermal emission can dominate the X-ray spectra. In some cases, the disk emission has been found to return to the disk and the reflection model assuming the primary continuum to be only a blackbody \citep{Garcia2022} fits the data well \citep[e.g.][]{Connors2020, Connors2021}. However, in the intermediate states, the strength of the disk and power-law components are comparable (see Fig.~\ref{resi} and Tab.~\ref{lp}). As a consequence, considering only one of the components (disk or corona) to ionize the disk atmosphere is not appropriate.

%\cite{Garcia2022} have computed a reflection model assuming the primary continuum to be a blackbody. Given the comparable strength of the disk and power-law components in the intermediate state (see Fig.~\ref{resi}), considering both contributions in ionizing the disk could be important.

Therefore, we implement the \texttt{refhiden} model that is specifically designed for accreting stellar mass black holes \citep{Ross2007}. In this model, the disk atmosphere is ionized by the combination of a power-law component from above and a blackbody from below. Note that here the blackbody is assumed to be single-temperature, which is different from the multi-temperature disk blackbody component. The model also includes the effect of Comptonization in the accretion disk. The full model reads as \texttt{tbabs*simplcut*relconv*refhiden} (Model 4). The broadening kernel \texttt{relconv} \citep{Dauser2010} is required to account for the relativistic effects. The disk thermal emission is already incorporated in \texttt{refhiden} so there is no need to add additional disk component. The parameters in the \texttt{refhiden} model include: the hydrogen number density on the disk ($H_{\rm den}$), the temperature of the blackbody from the disk ($kT_{\rm bb}$), the flux ratio between the power-law and blackbody components and  the power-law index $\Gamma$ (Illum/BB). We link $\Gamma$ of the \texttt{refhiden} to the same parameter in \texttt{simplcut}. To better constrain the evolution of other parameters, we fix the black hole spin at 0.998 for the fitting.

\begin{figure*}
    \centering
    \includegraphics[width=\linewidth]{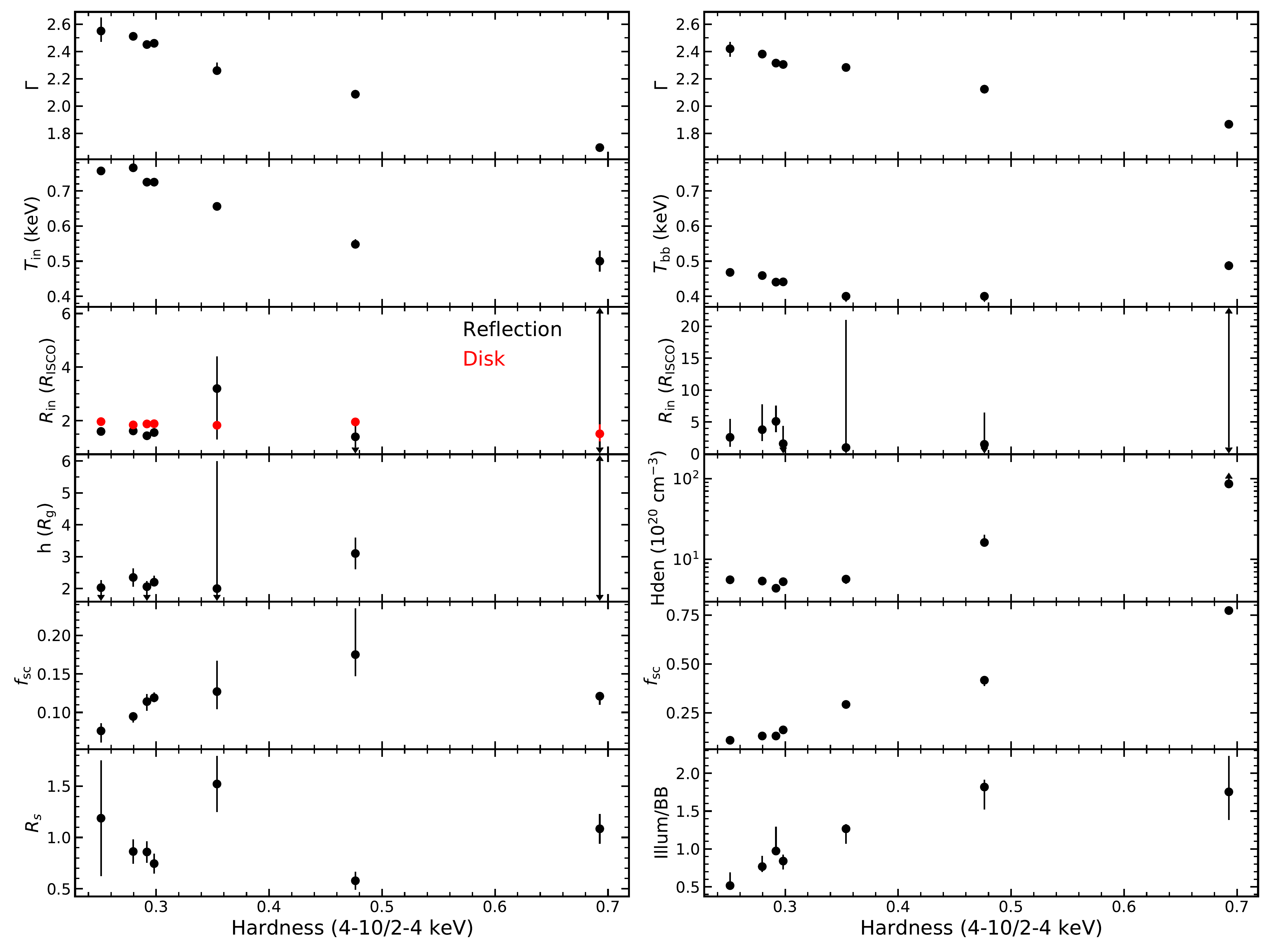}
    \caption{The evolution of spectral parameters along with X-ray hardness of the source. The definition of the hardness is the same as in Fig.~\ref{log}. Lower and upper limits are marked with arrows. Panels in the left show the parametes from Model 1A except the scattering fraction ($f_{\rm sc}$, from Model 3). In the third panel in the left, the red color represents estimation of the inner disk radius from the normalization parameter of \texttt{diskbb} of Model 3 and the size of the ISCO is chosen for a black hole with $a_*=0.95$. The reflection strength ($R_{\rm s}$) is defined as the ratio between the observed flux of the reflected and the direct corona emission in the 0.1--100 keV band. Parameters from Model 4 with the \texttt{refhiden} model are shown in the right panels.}
    \label{relation}
\end{figure*}

The model provides an acceptable fit ($\chi^2/\nu=3788.8/3386$) and the best-fit parameters are shown in Tab.~\ref{refhiden}. The variations of spectral parameters from this model during the state transition are shown in the right panels of Fig.~\ref{relation} and can be compared with that from Model 1A in the left panels. The \texttt{refhiden} model gives a blackbody temperature ($kT_{\rm bb}$) that is systematically lower than the $T_{\rm in}$ found in other fits with the \texttt{diskbb} model. This is within expectation because the temperature in \texttt{refhiden} is the effective mid-plane temperature of the disk. The disk surface temperature would be higher after including the effect of color correction \citep[see the discussion in][]{Reis2008}. Moreover, \texttt{diskbb} assumes a multi-temperature blackbody while a single temperature is used in \texttt{refhiden}. This is probably also affecting the value of the scattering fraction ($f_{\rm sc}$) of \texttt{simplcut} but the decreasing trend of this parameter along with the transition is retained. The number density of hydrogen is higher at the early phase of the transition (the first two epochs). One parameter of \texttt{refhiden} that clearly changes through the transition is the flux ratio between the illuminating power-law and the blackbody from the disk mid-plane.

% \begin{figure}
%     \centering
%     \includegraphics[width=\linewidth]{plots/eemodel_delchi.pdf}
%     \caption{The best-fit model for the seven epochs and the corresponding residuals.}
%     \label{resi}
% \end{figure}

\section{Discussion}\label{s-dis}

% Evolution of Tin, Rin, Gamma (Tau)

\subsection{Evolution of the disc-corona system}

For the transition from the hard to the soft state, \src{} gets progressively brighter in X-ray. In Fig.~\ref{relation}, we show how the spectral parameters evolve with X-ray hardness of the source. The state transition can be characterized by decreasing of the hardness. This process is apparently associated with the steepening of the corona emission component, with a photon index ($\Gamma$) evolving from 1.7 to 2.5 (see the top panel of Fig.~\ref{relation}). Meanwhile, the X-ray luminosity increases from 10\% to 20\% of the Eddington limit assuming a black hole mass of 10~$M_{\odot}$ and a distance of 8~kpc (see Tab.~\ref{relxillcp}). There is a clear trend of the inner disk temperature increasing from 0.5 to 0.76 keV. The contribution of the disk component (in 0.1--100 keV band) also increases from 13\% to 35\%. This increase of soft cooling photons could explain the softening of the corona emission \citep[e.g.][]{Del-Santo2008}.

As for the coronal temperature ($kT_{\rm e}$), the data do not provide good constraint on this parameter. Only lower limits are found for Epoch 2--7 (see Tab.~\ref{relxillcp}), while these lower limits are still higher than the measurement of Epoch 1. This weak tendency of increasing coronal temperature during the state transition is consistent with previous broadband spectral analysis of GX~339-4 \citep[e.g.][]{Motta2009MNRAS.400.1603M,Sridhar2020} and other systems \citep[e.g.][]{Zhang2022}. It is different from the tendency of decreasing coronal temperature with luminosity in the hard state \citep[e.g.][]{Garcia2015} which is thought to be a result of stronger cooling by the seed photons. What is happening in the state transition could be that the decrease of the optical depth \citep[e.g.][]{McConnell2002,Del-Santo2008} causes less scattering between the seed photons and corona electrons and then increases the coronal temperature. Note that this increase of coronal temperature will be limited by the run-away electron-positron pair production process in the corona \citep[e.g.][]{Fabian2015}.

Another parameter that is of interest is the inner radius of the accretion disk ($R_{\rm in}$). There are already reflection-based studies that have investigated the accretion geometry of \src{} in the hard state. In \cite{Garcia2015}, the luminosity range from 1.7\% to 17\% of the Eddington limit ($L_{\rm Edd}$) has been studied with data from \textit{RXTE}/PCA and the $R_{\rm in}$ is found to be close to the ISCO ($<5~R_{\rm ISCO}$). \cite{Plant2015} have studied similar luminosity levels with data from \textit{XMM-Newton} in the timing mode but found large truncations of the disk \citep[see also][]{Basak2016}. This discrepancy could be due to the complex pile-up effect in the timing mode that can affect the iron K line \citep[e.g.][]{Miller2010}. The regime below 1\% of the Eddington luminosity has also been studied with reflection models \citep[e.g.][]{Shidatsu2011, Petrucci2014, Wang2018, Wang2020}. These studies indicate that $R_{\rm in}$ should decrease with increasing luminosity and reaches 20~$R_{\rm g}$ when the luminosity is above 1\% Eddington \citep[see Fig.~6 of][]{Garcia2019}.

Our analysis concerns the reflection spectra of \src{} during the hard-to-soft transition that covers the luminosity range from 10\% to 20\% $L_{\rm Edd}$. The results show that $R_{\rm in}$ is not constrained in Epoch 1. This is probably because of the low statistics of the data since Epoch 1 has the shortest exposure among all epochs. After Epoch 2, $R_{\rm in}$ is consistent with the ISCO in some models and still below a few times the ISCO in other models. This indicates that the disk inner edge is relatively stable across the state transition.\footnote{We note that the \hxmt{} spectra of this outburst in the bright hard state ($>$ 8\% $L_{\rm Edd}$, before our Epoch 1) have also been analyzed and the inner radius is found to be constant ($\sim 3~R_{\rm ISCO}$, Ren et al. in prep).} The inner disc radius can also be estimated from the normalization parameter of the \texttt{diskbb} model \citep{Kubota1998}. We use values from Model 3 since it takes into account both the transmitted and scattered disk photons. The results are shown in Fig.~\ref{relation} in red assuming a distance of 8~kpc, a black hole mass of 10~$M_{\odot}$, an inclination of 40 deg, a color correction factor of 1.7 and a relativistic correction factor of 0.412. It again shows that the inner disk radius is stable through the transition although the absolute values need to be taken with caution as there are systematic uncertainties in the assumptions for mass, distance and color correction.

Similar results have been found by \cite{Sridhar2020} who studies the 2002--2003 and 2004--2005 outbursts of \src{} with data from \textit{RXTE}/PCA. In their work, by fixing the black hole spin at 0.998, the $R_{\rm in}$ is found to reach close to the ISCO at the onset of the transition for both outbursts even though the transition luminosity differs by a factor of 3. A stable disk during the state transition has also been found in other systems (e.g. XTE~J1550--564, \citealt{Rodriguez2003}; MAXI~J1820+070, \citealt{Wang2021}). %This suggests that the mechanisms that trigger the state transition should be attributed to the corona condition instead of the inner edge of the disk.

%In the bright hard state, other sources have also shown indication that the disk already extends to the ISCO (e.g. GRS~1739--278, \citealt{Miller2015}; XTE~J1752-223, \citealt{Garcia2018ApJ86425G}). 

%(\textit{Suzaku}), \cite{Wang2018} (\textit{Swift+NuSTAR})

The corona height shows similar tendency as $R_{\rm in}$, which is not constrained in Epoch 1 and remains close to the black hole ($<6~R_{\rm g}$) in the following observations. When the primary source is close to the black hole, the light bending effect would concentrate its radiation to the innermost part of the accretion disk \citep[e.g.][]{Dauser2013}. This is indeed what we find with the broken power-law emissivity, with which we obtain a steep inner emissivity index (see Tab.~\ref{bpo}). We also define an empirical reflection strength parameter ($R_{\rm s}$) which is the ratio between the reflected and the power-law flux in the 0.1--100 keV band. This is different from but similar to the reflection fraction parameter ($R_{\rm f}$), which is the ratio between the corona intensity that illuminates the disk and the corona intensity that reaches the observer \citep{Dauser2016}. Variation of the reflection strength may indicate changes of the disk-corona system \citep[e.g.][]{You2021}. The lowest panel in the left of Fig.~\ref{relation} shows no significant changes of this parameter, which is consistent with the prediction of a stable inner disk radius.

We note that relativistic reflection analysis based on the single lamppost geometry may only be sensitive to the parts of the corona that are close to the black hole since they dominate the reflection emissivity \citep[see][]{Wang2021}. In practice, the two-lamppost reflection models have been used to mimic a vertically extended corona \citep[e.g.][]{Buisson2019}. The upper lamppost could be responsible for the narrow cores in the iron band that is usually assumed to originate from a distant reflector \citep[e.g.][]{Garcia2019}. We do not explore further in this aspect since our data does not require the presence of a narrow iron line.

%\red{We also define an empirical reflection strength parameter ($R_{\rm s}$) which is the ratio between the observed flux of the reflected and the direct corona emission in the 20--40 keV band (see Sec.~\ref{comp}). From Fig.~\ref{relation}, we see that the reflection strength tentatively decreases from 1.5 to 0.5 during the transition. However, the large uncertainties do not}

The scattering fraction parameter ($f_{\rm sc}$) in \texttt{simplcut} is related to the geometry of the disk-corona system and also the optical depth of the corona. From Fig.~\ref{relation}, we see that, except for Epoch~1, $f_{\rm sc}$ from Model~3 shows a tentative decreasing trend through the transition although the uncertainties are large. The trend is more evident in Model~4 (see the right panel of Fig.~\ref{relation}). In the fit with \texttt{refhiden}, we find that the flux ratio between the illuminating power-law component and the blackbody component from the disk (Illum/BB) also decreases by a factor of 3 from Epoch~1 to Epoch~7. Meanwhile, the temperature of the disk mid-plane ($T_{\rm bb}$) only changes in a narrow range (0.4--0.5 keV). Given these variation patterns and the fact that the inner disk radius remains stable, we argue that the state transition is more likely to be related to the corona properties instead of the inner edge of the disk. This scenario is also supported by recent studies on the reverberation lag of black hole transients \citep{Wang2022ApJ...930...18W}, in which the lag is found to become longer across the hard-to-soft transition and is explained as a vertically expanding jet. Similar corona behavior has also been found in other sources by analyzing the lag spectrum with Comptonization models \citep[e.g.][]{Garcia2021MNRAS.501.3173G, Garcia2022MNRAS.513.4196G, Mendez2022NatAs...6..577M, Zhang2022,Peirano2023}.

The ionization parameter $\xi$ is defined as $\xi=L/nr^2$ where $L$ is the ionizing luminosity, $n$ is the density and $r$ is the distance to the illuminating source. We do not see strong variation on the ionization parameter. This is within our expectation since the source luminosity changes by only a factor of 2. The variation of the ionization parameter would be comparable to its fitting uncertainties if the density and distance do not change. Among the seven epochs, a high ionization parameter ($\log(\xi)>3.6$) is always required to fit the reflection spectra. This high ionization state is consistent with previous studies of \src{} at similar luminosity levels \citep[e.g.][]{Sridhar2020}. With the measured ionization parameter and the known source luminosity, we can calculate the disk density and compare it with our measurements (e.g. Tab.~\ref{relxilld}). Using the Equ. 9 of \cite{Zdziarski2020}, we calculate a disk density of $n_{\rm cal}\sim10^{22}$~cm$^{-3}$ with the averaged illuminating flux and ionization parameter in Tab.~\ref{relxilld}. This value is higher than what we measured with the \texttt{relxillD} model and exceeds the upper limit ($10^{19}$~cm$^{-3}$)\footnote{Note that in the latest version of \texttt{relxill}, which was released after writing of this paper, the density is allowed to vary up to $10^{20}$~cm$^{-3}$. See \url{http://www.sternwarte.uni-erlangen.de/~dauser/research/relxill/}.} allowed by the model. Similar discrepancy has been found in other intermediate data of \src{} \citep[e.g.][]{Zdziarski2020}. It might be explained if the inner disk radius is 30 times larger than the ISCO but it would conflict with the broad iron line we are seeing. More work in developing the reflection models would help to better understand this problem.

% We also define an empirical reflection strength parameter ($R_{\rm s}$) which is the ratio between the observed flux of the reflected and the direct corona emission in the 0.1--100 keV band.

% Parameters: spin, inclination, iron abundance systematic uncertainties (emissivity)

\subsection{Measurement of system parameters}
\label{measure}

\begin{figure*}[htbp]
    \centering
    \includegraphics[width=0.48\linewidth]{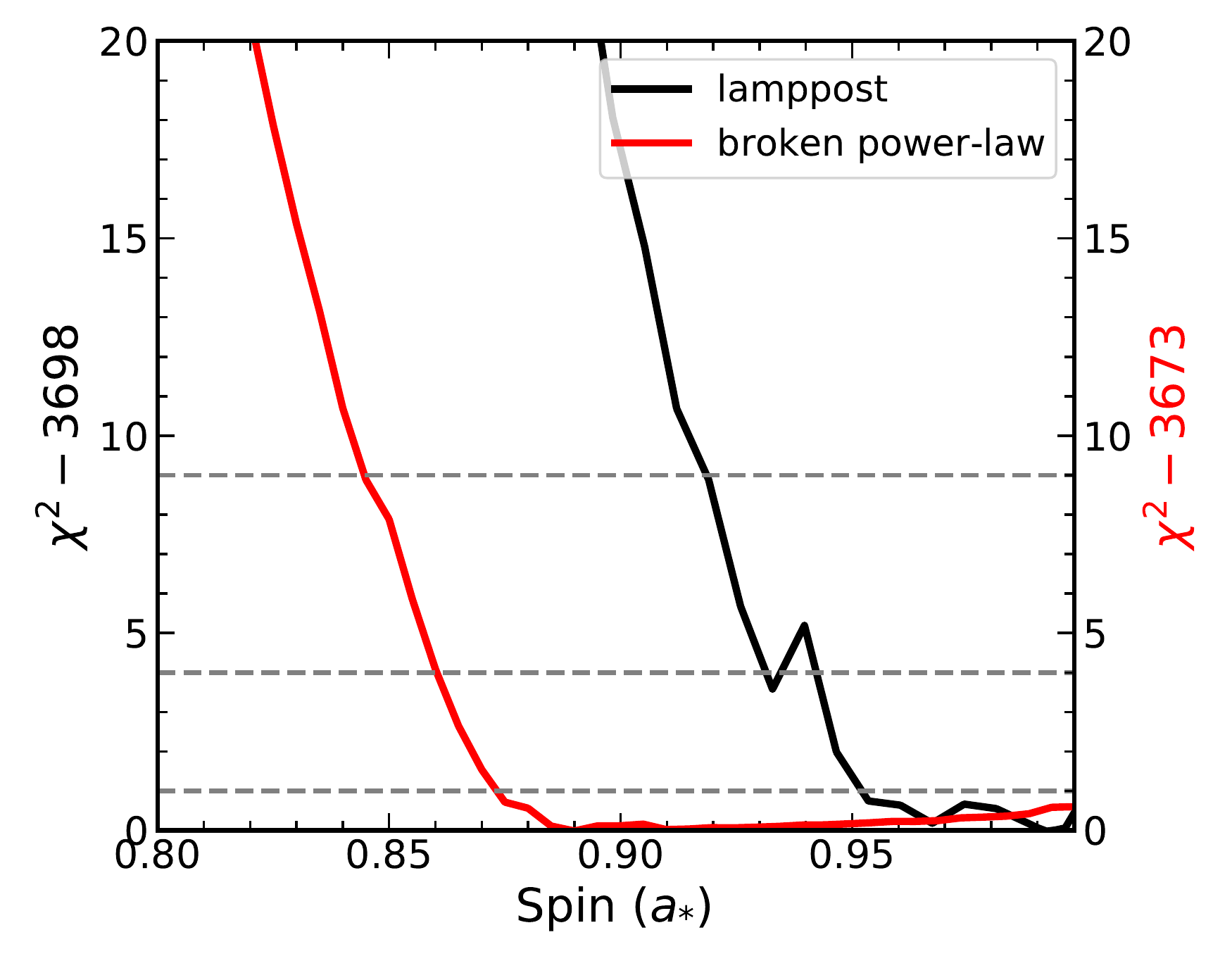}
    \includegraphics[width=0.48\linewidth]{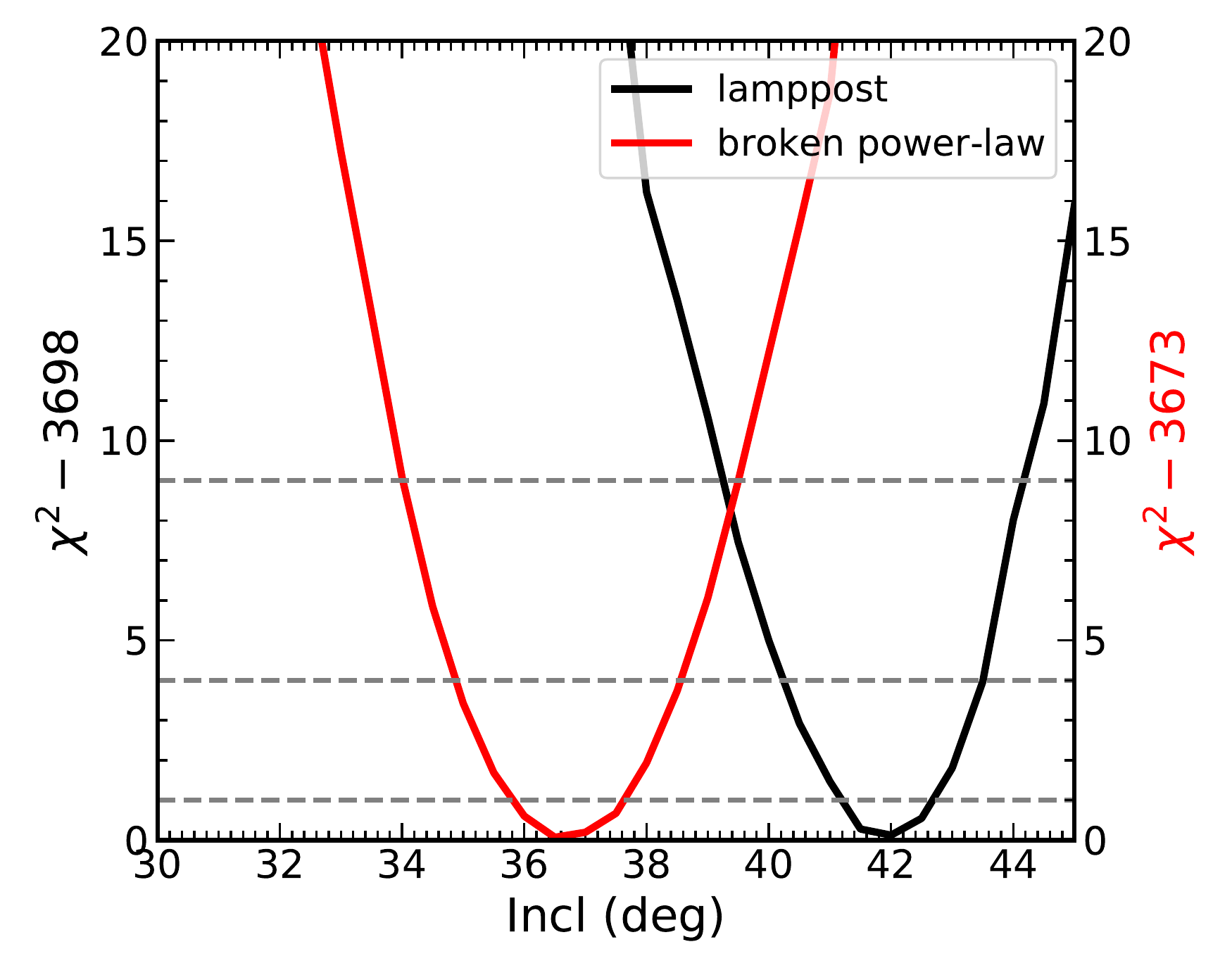}\\
    \caption{The $\chi^2$ contours for the black hole spin and the inclination parameters for the lamppost geometry and broken power-law emissivity. The horizontal lines represent the 1$\sigma$, 2$\sigma$ and 3$\sigma$ confidence levels for a single parameter of interest.}
    \label{del}
\end{figure*}

By fitting the seven spectra simultaneously with relativistic reflection models, we are able to measure the black hole spin parameter of \src{}. We obtain a constraint of $a_*>0.95$ if assuming a lamppost geometry. When implementing a broken power-law emissivity in \texttt{relxill}, we obtain the lowest $\chi^2$ among all models and get a constraint of $a_*>0.86$ (see Fig.~\ref{del}). This measurement confirms the high spin nature of the black hole in \src{} as found in previous studies with data in the hard \citep[e.g.][]{Reis2008, Ludlam2015, Garcia2015} or soft states \citep[e.g.][]{Parker2016}. 

Previous reflection based analysis have constrained the disk inclination angle of \src{} to be 30$^\circ$--60$^\circ$ \citep[e.g.][]{Garcia2015, Parker2016}. Moreover, by studying the near-infrared lightcurve of \src{}, \cite{Heida2017} have measured the orbital inclination to be 37$^\circ$--78$^\circ$. A more recent work by \cite{Zdziarski2019} updated the inclination to $\approx$ 40$^\circ$--60$^\circ$ based on considering evolutionary models for the donor. Our measurement of the inclination angle of the inner accretion disk ranges from 35 to 43 degree, which is consistent with the previous studies via different methods. We also find that the constraint on the inclination with reflection spectroscopy slightly changes ($42.0_{-1.4}^{+1.3}$ with the lamppost geometry and $36.6_{-1.3}^{+1.5}$ for the broken power-law emissivity) with the treatments of the emissivity profile (see Fig.~\ref{del}). These differences on parameter measurements due to systematic uncertainties are expected for reflection spectroscopy \citep[e.g.][]{Zhou2020PhRvD.101d3010Z,2020PhRvD.101l3014C}. It is a result of the lack of our knowledge about the system. We note that these systematic uncertainties should be considered when reporting measurements of parameters with disk reflection models. Other sources of systematic uncertainties include simplifications in the models \citep[e.g. the discussion in][]{Tripathi2021ApJ...913...79T} or instrumental effects \citep[see the review in][]{Bambi2021}.

%The result also shows that the data quality of \hxmt{} is capable of constraining the spin value of black holes in XRBs with relativistic reflection features.

We always obtain a very high iron abundance that is near ten times of the solar value. We note that the measurement may not represent the true abundance of the system. The super-solar iron abundance is a well known issue in the fitting of the relativistic reflection features \citep[e.g.][]{Dauser2012, Jiang2018, Garcia2018ASPC}. It has been shown that allowing the electron density ($n_{\rm e}$) of the disk to go higher than what is assumed in the standard reflection model ($\log(n_{\rm e}/{\rm cm}^{-3})$=15) could help to lower the iron abundance \citep[e.g.][]{Tomsick2018, Jiang2019b}. However, our high-density model still gives a very high iron abundance ($A_{\rm Fe}>8.6$). Another solution to solve this high iron abundance problem is to allow different photon indexes for the directly observed primary emission and the reflection spectrum \citep[e.g.][]{Furst2015}. We test this scenario but it does not help to lower the iron abundance (with $A_{\rm Fe}>9.0$). We note that the corona structure may be more complex than two uniform clouds of plasma \citep[e.g.][]{Uttley2011}. Moreover, \cite{Ross2002} have found that ignoring the returning radiation could also affect the measurement of the iron abundance while more recent studies on the returning radiation do not find strong impact \citep[e.g.][]{Riaz2021, Dauser2022}. There is also the possibility that the high iron abundance is real and caused by radiative levitation of metal ions in the inner part of the accretion disk \cite[e.g.][]{Reynolds2012}.

%Further developments in modeling and observation are required to solve this issue.

%%%%%%%%%%%%%%%%%%%%%%%%%%%%%%%

\section{Conclusion}\label{s-con}
In this work, we analyzed the broadband spectra (2--100 keV) of \src{} during the bright state transition of its 2021 outburst observed by \hxmt{}. Strong relativistic reflection features were found in the spectra, which allowed us to study the evolution of the disk-corona system during the transition. The main results are the following:

\begin{itemize}
    \item The hard to soft transition is associated with a stronger contribution of the disk thermal component to the X-ray spectrum and a steepening of the power-law emission.
    \item The inner disk radius stays close to the ISCO during the transition. This is even true when we consider a reflection model that assumes the ionzing continuum to be a combination of the corona emission from above and a blackbody emission from below.
    \item If assuming a lamppost geometry, the mesured corona height is always close to the black hole.
    \item The scattering fraction, which is the fraction of the disk photons that are scattered by the hot corona, decreases along with the transition.
    \item The data provide constraint on the black hole spin ($a_*>0.86$) and the inclination parameter ($i \approx$ 35$^\circ$--43$^\circ$) of the system.
\end{itemize}

{\bf Acknowledgments --} We thank the anonymous referee for helpful comments to improve the manuscript. We thank Yuexin Zhang for useful discussion. This work was supported by the National Natural Science Foundation of China (NSFC), Grant No. 11973019 and Grant No. 12250610185, the Natural Science Foundation of Shanghai, Grant No. 22ZR1403400, the Shanghai Municipal Education Commission, Grant No. 2019-01-07-00-07-E00035, and Fudan University, Grant No. JIH1512604. JJ acknowledges support from the Leverhulme Trust, the Isaac Newton Trust and St Edmund's College.

%%%%%%%%%%%%%%%%%%%%%%%%%%%%%%%

%%%%%%%%%% REFERENCE %%%%%%%%%

\bibliographystyle{apj}
\bibliography{bibliography}

%%%%%%%%%%%%%%%%%%%%%%%%%%%%%%%

\appendix

\section{The HXMT observations}
% observation ID
\begin{table}
    \centering
    \renewcommand\arraystretch{2.0}
    \caption{\hxmt{} observations of \src{} analyzed in this paper}
    \label{info-obs}
    %\begin{tabular*}{0.45\textwidth}{@{\extracolsep{\fill}}cccc}
    \begin{tabular}{cccccccc}
        \hline\hline
        Reference name & Date$^1$ & obsID & \multicolumn{3}{c}{Exposure (ks)} & Start time & Stop time\\
        & & & LE & ME & HE & MJD & MJD \\
        \hline
        Epoch 1 & 20210325 & P030402403702 & 1.86 & 3.31 &  2.30 & 59298.012 & 59298.151  \\ 
        & & P030402403703 & &  & &  59298.151 & 59298.289 \\
        \hline
        Epoch 2 & 20210326 & P030402403801 & 3.46 & 5.75 & 7.95 & 59299.602 & 59299.776 \\
        & & P030402403802 & &  & & 59299.776 & 59299.909 \\
        & & P030402403803 & &  & & 59299.909 & 59300.003 \\
        \hline
        Epoch 3 & 20210327 & P030402403901 & 2.40 & 3.86 & 3.88 & 59300.796 & 59300.923 \\
        &  & P030402403902 &  &  & & 59300.923 & 59301.059 \\
        \hline
        Epoch 4 & 20210328 & P030402403903 & 7.25 & 12.9 & 13.2 & 59301.059 & 59301.204 \\
        & & P030402403904 &  &  & & 59301.204 & 59301.343 \\
        & & P030402403905 &  &  & & 59301.343 & 59301.498 \\
        & & P030402403906 &  &  & & 59301.498 & 59301.634 \\
        & & P030402403907 &  &  & & 59301.634 & 59301.794 \\
        & & P030402403908 &  &  & & 59301.794 & 59301.916 \\
        & & P030402403909 &  &  & & 59301.916 & 59302.052 \\
        \hline
        Epoch 5 & 20210329 & P030402403910 & 7.46 & 13.3 & 13.6 & 59302.052 & 59302.189 \\
        & & P030402403911 &  &  & & 59302.189 & 59302.347 \\
        & & P030402403912 &  &  & & 59302.347 & 59302.515 \\
        & & P030402403913 &  &  & & 59302.515 & 59302.648 \\
        & & P030402403914 &  &  & & 59302.648 & 59302.787 \\
        & & P030402403915 &  &  & & 59302.787 & 59302.961 \\
        & & P030402403916 &  &  & & 59302.961 & 59303.120 \\
        \hline
        Epoch 6 & 20210330 & P030402403917 & 5.63 & 11.2 & 11.4 & 59303.120 & 59303.260 \\
        & & P030402403918 &  &  & & 59303.260 & 59303.399 \\
        & & P030402403919 &  &  & & 59303.399 & 59303.551 \\
        & & P030402403920 &  &  & & 59303.551 & 59303.684 \\
        & & P030402403921 &  &  & & 59303.684 & 59303.816 \\
        & & P030402403922 &  &  & & 59303.816 & 59303.949 \\
        & & P030402403923 &  &  & & 59303.949 & 59304.081 \\
        \hline
        Epoch 7 & 20210331 & P030402403924 & 4.57 & 9.95 & 10.2 & 59304.081 & 59304.214 \\
        & & P030402403925 &  &  & & 59304.214 & 59304.346 \\
        & & P030402403926 &  &  & & 59304.346 & 59304.479 \\
        & & P030402403927 &  &  & & 59304.479 & 59304.611 \\
        & & P030402403928 &  &  & & 59304.611 & 59304.743 \\
        & & P030402403929 &  &  & & 59304.743 & 59304.876 \\
        & & P030402403930 &  &  & & 59304.876 & 59305.037 \\
        \hline

    \end{tabular}\\

\textit{Note}. (1) The observation date is presented in the form of \texttt{yyyymmdd}.
\end{table}
% Only LE exposures are listed for \hxmt{} observations.

\section{The best-fit tables}

\begin{table*}
    \centering
    \caption{Best-fit values with a broken power-law emissivity with $q_{\rm out}=3$ (Model 1B).}
    \label{bpo}
    \renewcommand\arraystretch{2.0}
    \begin{tabular}{lcccccccc}
        \hline\hline
        Component       & Parameter & Epoch 1 & Epoch 2 & Epoch 3 & Epoch 4 & Epoch 5 & Epoch 6 & Epoch 7 \\
        \hline
        \texttt{tbabs} & $N_{\rm H}$ (10$^{22}$ cm$^{-2}$) & \multicolumn{7}{c}{$0.99_{-0.06}^{+0.05}$} \\
        \hline
        \texttt{diskbb} & $T_{\rm in}$ (keV) & $0.50_{-0.04}^{+0.04}$ & $0.540_{-0.012}^{+0.013}$ & $0.658_{-0.009}^{+0.009}$ & $0.718_{-0.011}^{+0.008}$ & $0.726_{-0.011}^{+0.01}$ & $0.765_{-0.008}^{+0.009}$ & $0.752_{-0.009}^{+0.011}$\\
        \hline
        \texttt{relxill} & $a_*$ & \multicolumn{7}{c}{$>0.86$} \\
        & $q_{\rm in}$ & $3^{*}$ & $3^{*}$ & $3^{*}$ & $10_{-3}^{+P}$ & $10_{-4}^{+P}$ & $9_{-4}^{+P}$ & $7.4_{-1.7}^{+P}$\\
        & $q_{\rm out}$ & \multicolumn{7}{c}{$3^{*}$} \\
        & $R_{\rm br}$ ($R_{\rm g}$) & - & - & - & $3.5_{-0.3}^{+0.5}$ & $3.6_{-0.3}^{+0.8}$ & $3.62_{-0.28}^{+1.3}$ & $4.2_{-0.9}^{+0.9}$\\
        & Incl & \multicolumn{7}{c}{$36.6_{-1.3}^{+1.5}$} \\
        & $A_{\rm Fe}$ (solar) & \multicolumn{7}{c}{$10.0_{-1.1}^{+P}$} \\
        & $R_{\rm in}$ (ISCO) & $40_{-33}^{+P}$ & $3.6_{-1.6}^{+4.0}$ & $2.3_{-0.6}^{+1.4}$ & $1.0_{-P}^{+0.8}$ & $1.0_{-P}^{+0.6}$ & $1.1_{-P}^{+0.9}$ & $1.0_{-P}^{+0.7}$\\
        & $\log(\xi)$ & $4.29_{-0.1}^{+0.05}$ & $4.54_{-0.07}^{+0.05}$ & $4.60_{-0.1}^{+0.07}$ & $4.02_{-0.15}^{+0.21}$ & $4.35_{-0.22}^{+0.12}$ & $4.31_{-0.16}^{+0.14}$ & $4.1_{-0.3}^{+0.3}$\\
        \hline
        \texttt{cutoffpl} & $\Gamma$ & $1.701_{-0.025}^{+0.029}$ & $2.054_{-0.013}^{+0.017}$ & $2.27_{-0.06}^{+0.04}$ & $2.42_{-0.04}^{+0.04}$ & $2.42_{-0.04}^{+0.03}$ & $2.504_{-0.027}^{+0.027}$ & $2.48_{-0.1}^{+0.15}$\\
        & $E_{\rm cut}$ (keV) & $140_{-20}^{+30}$ & $>450$ & $>400$ & $>220$ & $>330$ & $>410$ & $130_{-60}^{+70}$\\
        \hline
        \texttt{cflux} \\
        $F_{\rm diskbb}$ & (10$^{-8}$~erg~cm$^{-2}$~s$^{-1}$) & $0.219_{-0.024}^{+0.03}$ & $0.556_{-0.029}^{+0.020}$ & $0.93_{-0.03}^{+0.04}$ & $1.310_{-0.025}^{+0.023}$ & $1.360_{-0.023}^{+0.03}$ & $1.597_{-0.025}^{+0.022}$ & $1.85_{-0.06}^{+0.04}$\\
        $F_{\rm relxill}$ & (10$^{-8}$~erg~cm$^{-2}$~s$^{-1}$) & $0.73_{-0.13}^{+0.16}$ & $1.26_{-0.08}^{+0.07}$ & $1.44_{-0.19}^{+0.17}$ & $1.38_{-0.12}^{+0.14}$ & $1.45_{-0.18}^{+0.11}$ & $1.62_{-0.16}^{+0.2}$ & $1.56_{-0.29}^{+0.5}$\\
        $F_{\rm cutoffpl}$ & (10$^{-8}$~erg~cm$^{-2}$~s$^{-1}$) & $0.73_{-0.20}^{+0.07}$ & $0.57_{-0.12}^{+0.15}$ & $0.9_{-0.3}^{+0.5}$ & $1.78_{-0.24}^{+0.25}$ & $1.52_{-0.25}^{+0.22}$ & $1.80_{-0.4}^{+0.23}$ & $1.3_{-0.4}^{+0.6}$\\
        \hline
        \texttt{constant} & ME/LE & \multicolumn{7}{c}{$1.035_{-0.014}^{+0.014}$} \\
        & HE/LE & \multicolumn{7}{c}{$1.158_{-0.023}^{+0.018}$} \\ 
        \hline
                        & $\chi^2$/d.o.f &  \multicolumn{7}{c}{3673.1/3383} \\
        \hline\hline
    \end{tabular}
    \textit{Note.} Best-fit parameters for the model \texttt{tbabs*(diskbb+relxill+cutoffpl)}. Parameters with $^*$ are fixed during the fit and the symbol $P$ denotes the upper or lower boundary. Parameters that are only shown for Epoch~4 are tied across all observations.
\end{table*}

\begin{table*}
    \centering
    \caption{Best-fit values with \texttt{relxillD}.}
    \label{relxilld}
    \renewcommand\arraystretch{2.0}
    \begin{tabular}{lcccccccc}
        \hline\hline
        Component       & Parameter & Epoch 1 & Epoch 2 & Epoch 3 & Epoch 4 & Epoch 5 & Epoch 6 & Epoch 7 \\
        \hline
        \textsc{tbabs} & $N_{\rm H}$ (10$^{22}$ cm$^{-2}$) & \multicolumn{7}{c}{$0.98_{-0.04}^{+0.04}$} \\
        \hline
        \textsc{diskbb} & $T_{\rm in}$ (keV) & $0.48_{-0.04}^{+0.04}$ & $0.561_{-0.014}^{+0.012}$ & $0.659_{-0.008}^{+0.008}$ & $0.717_{-0.01}^{+0.008}$ & $0.726_{-0.007}^{+0.006}$ & $0.762_{-0.007}^{+0.008}$ & $0.756_{-0.006}^{+0.008}$\\
        %& Norm & $1850_{-600}^{+1100}$ & $2660_{-260}^{+360}$ & $2400_{-180}^{+150}$ & $2400_{-120}^{+220}$ & $2400_{-120}^{+110}$ & $2300_{-100}^{+100}$ & $2600_{-100}^{+160}$\\
        \hline
        \textsc{relxill} & $a_*$ & \multicolumn{7}{c}{$>0.87$} \\
        & $q_{\rm in}$ & $3^{*}$ & $3^{*}$ & $3^{*}$ & $7.9_{-4}^{+P}$ & $7.3_{-2.2}^{+P}$ & $7.2_{-2.7}^{+P}$ & $6.5_{-1.4}^{+P}$\\
        & $q_{\rm out}$ & \multicolumn{7}{c}{$3^{*}$} \\
        & $R_{\rm br}$ ($R_{\rm g}$) & - & - & - & $3.7_{-0.4}^{+0.7}$ & $3.8_{-0.6}^{+1.0}$ & $3.8_{-0.6}^{+1.2}$ & $4.2_{-0.8}^{+1.1}$\\
        & Incl & \multicolumn{7}{c}{$35.3_{-1.5}^{+1.5}$} \\
        & $A_{\rm Fe}$ & \multicolumn{7}{c}{$>8.6$} \\
        & $R_{\rm in}$ (ISCO) & $11_{-6}^{+P}$ & $1.5_{-P}^{+0.8}$ & $1.5_{-P}^{+0.9}$ & $1.69_{-0.23}^{+0.22}$ & $1.7_{-0.4}^{+0.3}$ & $1.77_{-0.26}^{+0.29}$ & $1.73_{-0.18}^{+0.25}$\\
        & $\log(\xi)$ & $3.97_{-0.19}^{+0.17}$ & $3.69_{-0.14}^{+0.17}$ & $4.2_{-0.1}^{+0.1}$ & $3.49_{-0.15}^{+0.13}$ & $4.00_{-0.16}^{+0.22}$ & $4.0_{-0.4}^{+0.2}$ & $4.01_{-0.4}^{+0.27}$\\
        & $\log(n_{\rm e})$ & $16_{-P}^{+P}$ & $19.0_{-0.6}^{+P}$ & $19.0_{-0.6}^{+P}$ & $18.9_{-0.9}^{+P}$ & $17.9_{-1.3}^{+0.4}$ & $18.2_{-0.5}^{+0.5}$ & $15.0_{-P}^{+2.1}$\\
        %& Norm & $0.0062_{-0.0013}^{+0.001}$ & $0.0073_{-0.0006}^{+0.0014}$ & $0.0118_{-0.0013}^{+0.0017}$ & $0.0163_{-0.0009}^{+0.0026}$ & $0.0155_{-0.0017}^{+0.0028}$ & $0.0156_{-0.0018}^{+0.003}$ & $0.025_{-0.004}^{+0.013}$\\
        \hline
        \textsc{cutoffpl} & $\Gamma$ & $1.724_{-0.019}^{+0.026}$ & $2.042_{-0.011}^{+0.024}$ & $2.166_{-0.022}^{+0.04}$ & $2.369_{-0.029}^{+0.019}$ & $2.387_{-0.024}^{+0.023}$ & $2.43_{-0.04}^{+0.04}$ & $2.54_{-0.06}^{+0.09}$\\
        & $E_{\rm cut}$ (keV) & $100_{-14}^{+18}$ & $130_{-20}^{+30}$ & $>270$ & $>320$ & $>310$ & $>390$ & $170_{-90}^{+270}$\\
        \hline
        \texttt{cflux} \\
        $F_{\rm diskbb}$ & (10$^{-8}$~erg~cm$^{-2}$~s$^{-1}$) & $0.22_{-0.03}^{+0.04}$ & $0.556_{-0.029}^{+0.027}$ & $0.965_{-0.017}^{+0.017}$ & $1.346_{-0.012}^{+0.028}$ & $1.390_{-0.017}^{+0.027}$ & $1.634_{-0.012}^{+0.029}$ & $1.82_{-0.03}^{+0.05}$\\
        $F_{\rm relxillD}$ & (10$^{-8}$~erg~cm$^{-2}$~s$^{-1}$) & $0.55_{-0.08}^{+0.09}$ & $0.69_{-0.1}^{+0.04}$ & $1.2_{-0.1}^{+0.1}$ & $1.43_{-0.3}^{+0.11}$ & $1.28_{-0.11}^{+0.13}$ & $1.44_{-0.18}^{+0.3}$ & $1.73_{-0.22}^{+0.21}$\\
        $F_{\rm cutoffpl}$ & (10$^{-8}$~erg~cm$^{-2}$~s$^{-1}$) & $1.00_{-0.05}^{+0.09}$ & $1.21_{-0.06}^{+0.11}$ & $0.75_{-0.19}^{+0.07}$ & $1.54_{-0.14}^{+0.1}$ & $1.53_{-0.11}^{+0.08}$ & $1.51_{-0.3}^{+0.11}$ & $1.6_{-0.6}^{+0.6}$\\
        %& Norm & $0.70_{-0.19}^{+0.06}$ & $1.29_{-0.08}^{+0.12}$ & $0.80_{-0.08}^{+0.26}$ & $1.67_{-0.16}^{+0.11}$ & $1.62_{-0.12}^{+0.17}$ & $1.58_{-0.3}^{+0.19}$ & $1.6_{-0.4}^{+0.4}$\\
        \hline
        \texttt{constant} & ME/LE & \multicolumn{7}{c}{$1.025_{-0.014}^{+0.016}$} \\
        & HE/LE & \multicolumn{7}{c}{$1.088_{-0.021}^{+0.022}$} \\ 
        \hline
                        & $\chi^2$/d.o.f &  \multicolumn{7}{c}{3664.5/3376} \\
        \hline\hline
    \end{tabular}
    \textit{Note.} Best-fit parameters for the model \texttt{tbabs*(diskbb+relxillD+cutoffpl)}. Parameters with $^*$ are fixed during the fit and the symbol $P$ denotes the upper or lower boundary. Parameters that are only shown for Epoch~4 are tied across all observations.
\end{table*}

\begin{table*}
    \centering
    \caption{Best-fit values with \texttt{relxillcp} (Model 2).}
    \label{relxillcp}
    \renewcommand\arraystretch{2.0}
    \begin{tabular}{lcccccccc}
        \hline\hline
        Component       & Parameter & Epoch 1 & Epoch 2 & Epoch 3 & Epoch 4 & Epoch 5 & Epoch 6 & Epoch 7 \\
        \hline
        \texttt{tbabs} & $N_{\rm H}$ (10$^{22}$ cm$^{-2}$) & \multicolumn{7}{c}{$0.73_{-0.06}^{+0.06}$} \\
        \hline
        \texttt{diskbb} & $T_{\rm in}$ (keV) & $0.48_{-0.04}^{+0.04}$ & $0.567_{-0.016}^{+0.016}$ & $0.676_{-0.01}^{+0.009}$ & $0.729_{-0.009}^{+0.008}$ & $0.739_{-0.009}^{+0.008}$ & $0.775_{-0.009}^{+0.008}$ & $0.768_{-0.009}^{+0.008}$\\
        \hline
        \texttt{relxillcp} & $a_*$ & \multicolumn{7}{c}{$>0.85$} \\
        & $q_{\rm in}$ & $3^{*}$ & $3^{*}$ & $3^{*}$ & $10_{-3}^{+P}$ & $8.0_{-2.2}^{+P}$ & $9_{-4}^{+P}$ & $7.3_{-2.0}^{+P}$\\
        & $q_{\rm out}$ & \multicolumn{7}{c}{$3^{*}$} \\
        & $R_{\rm br}$ ($R_{\rm g}$) & - & - & - & $3.42_{-0.12}^{+0.6}$ & $3.9_{-0.7}^{+0.9}$ & $3.6_{-0.3}^{+2.1}$ & $4.1_{-0.9}^{+1.2}$\\
        & Incl & \multicolumn{7}{c}{$37.1_{-1.3}^{+1.4}$} \\
        & $A_{\rm Fe}$ & \multicolumn{7}{c}{$10.0_{-0.3}^{+P}$} \\
        & $R_{\rm in}$ (ISCO) & $>12$ & $2.3_{-0.8}^{+1.8}$ & $2.5_{-1.1}^{+1.8}$ & $1.1_{-P}^{+0.7}$ & $1.1_{-P}^{+0.4}$ & $1.18_{-0.12}^{+0.5}$ & $1.1_{-P}^{+0.6}$\\
        & $\log(\xi)$ & $4.56_{-0.11}^{+0.05}$ & $4.46_{-0.1}^{+0.13}$ & $4.66_{-0.07}^{+P}$ & $3.99_{-0.15}^{+0.12}$ & $4.09_{-0.18}^{+0.24}$ & $4.27_{-0.24}^{+0.12}$ & $3.93_{-0.3}^{+0.22}$\\
        \hline
        \texttt{nthcomp} & $\Gamma$ & $1.773_{-0.024}^{+0.04}$ & $2.092_{-0.022}^{+0.019}$ & $2.215_{-0.027}^{+0.03}$ & $2.339_{-0.016}^{+0.02}$ & $2.360_{-0.023}^{+0.022}$ & $2.391_{-0.018}^{+0.024}$ & $2.42_{-0.08}^{+0.12}$\\
        & $kT_{\rm e}$ (keV) & $64_{-21}^{+30}$ & $160_{-70}^{+600}$ & $>270$ & $>400$ & $>310$ & $>410$ & $65_{-30}^{+600}$\\
        \hline
        \texttt{cflux} \\
        $F_{\rm diskbb}$ & (10$^{-8}$~erg~cm$^{-2}$~s$^{-1}$) & $0.183_{-0.024}^{+0.03}$ & $0.506_{-0.022}^{+0.024}$ & $0.938_{-0.012}^{+0.025}$ & $1.370_{-0.022}^{+0.022}$ & $1.421_{-0.022}^{+0.024}$ & $1.665_{-0.022}^{+0.023}$ & $1.86_{-0.03}^{+0.03}$\\
        $F_{\rm relxillcp}$ & (10$^{-8}$~erg~cm$^{-2}$~s$^{-1}$) & $0.94_{-0.25}^{+0.15}$ & $0.85_{-0.15}^{+0.12}$ & $1.10_{-0.14}^{+0.12}$ & $1.04_{-0.07}^{+0.09}$ & $1.05_{-0.07}^{+0.09}$ & $1.09_{-0.09}^{+0.11}$ & $1.11_{-0.19}^{+0.6}$\\
        $F_{\rm nthcomp}$ & (10$^{-8}$~erg~cm$^{-2}$~s$^{-1}$) & $0.50_{-0.15}^{+0.24}$ & $0.67_{-0.08}^{+0.11}$ & $0.46_{-0.08}^{+0.08}$ & $0.528_{-0.027}^{+0.022}$ & $0.50_{-0.03}^{+0.03}$ & $0.43_{-0.03}^{+0.03}$ & $0.34_{-0.06}^{+0.04}$\\
        \hline
        & $L_{\rm 0.1-100 keV}/L_{\rm Edd}$ (\%) & 10.4 & 12.9 & 15.9 & 18.8 & 19.0 & 20.3 & 21.1 \\
        \hline
        \texttt{constant} & ME/LE & \multicolumn{7}{c}{$1.022_{-0.012}^{+0.014}$} \\
        & HE/LE & \multicolumn{7}{c}{$1.147_{-0.025}^{+0.034}$} \\ 
        \hline
                        & $\chi^2$/d.o.f &  \multicolumn{7}{c}{3737.3/3383} \\
        \hline\hline
    \end{tabular}
    \textit{Note.} Best-fit parameters for the model \texttt{tbabs*(diskbb+relxillcp+cutoffpl)}. Parameters with $^*$ are fixed during the fit and the symbol $P$ denotes the upper or lower boundary. Parameters that are only shown for Epoch~4 are tied across all observations.
\end{table*}

\begin{table*}
    \centering
    \caption{Best-fit values with \texttt{simplcut} (Model 3).}
    \label{simplcut}
    \renewcommand\arraystretch{2.0}
    \begin{tabular}{lcccccccc}
        \hline\hline
        Component       & Parameter & Epoch 1 & Epoch 2 & Epoch 3 & Epoch 4 & Epoch 5 & Epoch 6 & Epoch 7 \\
        \hline
        \texttt{tbabs} & $N_{\rm H}$ (10$^{22}$ cm$^{-2}$) & \multicolumn{7}{c}{$0.75_{-0.06}^{+0.04}$} \\
        \hline
        \texttt{diskbb} & $T_{\rm in}$ (keV) & $0.50_{-0.04}^{+0.04}$ & $0.575_{-0.017}^{+0.018}$ & $0.682_{-0.009}^{+0.008}$ & $0.735_{-0.007}^{+0.007}$ & $0.743_{-0.01}^{+0.007}$ & $0.777_{-0.007}^{+0.007}$ & $0.769_{-0.009}^{+0.008}$\\
        \hline
        \texttt{relxillcp} & $a_*$ & \multicolumn{7}{c}{$0.92_{-0.07}^{+0.05}$} \\
        & $q_{\rm in}$ & $3^{*}$ & $3^{*}$ & $3^{*}$ & $9.7_{-3}^{+P}$ & $7.8_{-2.3}^{+P}$ & $8.9_{-5}^{+P}$ & $7.0_{-2.0}^{+P}$\\
        & $q_{\rm out}$ & \multicolumn{7}{c}{$3^{*}$} \\
        & $R_{\rm br}$ ($R_{\rm g}$) & - & - & - & $3.32_{-0.15}^{+0.5}$ & $3.8_{-0.4}^{+1}$ & $3.45_{-0.28}^{+2.6}$ & $4.0_{-0.7}^{+1.5}$\\
        & Incl & \multicolumn{7}{c}{$37.3_{-1.1}^{+1.4}$} \\
        & $A_{\rm Fe}$ & \multicolumn{7}{c}{$10.0_{-0.3}^{+P}$} \\
        & $R_{\rm in}$ (ISCO) & $50.0_{-39}^{+P}$ & $3.0_{-0.9}^{+1.9}$ & $3.1_{-0.9}^{+1.0}$ & $1.4_{-P}^{+0.5}$ & $1.41_{-0.30}^{+0.25}$ & $1.47_{-0.29}^{+0.24}$ & $1.4_{-0.3}^{+0.5}$\\
        & $\log(\xi)$ & $4.62_{-0.04}^{+0.03}$ & $4.59_{-0.15}^{+0.09}$ & $4.70_{-0.19}^{+P}$ & $3.99_{-0.16}^{+0.12}$ & $4.13_{-0.22}^{+0.22}$ & $4.28_{-0.24}^{+0.13}$ & $3.94_{-0.27}^{+0.6}$\\
        & $\Gamma$ & $1.771_{-0.012}^{+0.008}$ & $2.125_{-0.022}^{+0.024}$ & $2.243_{-0.029}^{+0.04}$ & $2.356_{-0.018}^{+0.03}$ & $2.370_{-0.022}^{+0.019}$ & $2.397_{-0.016}^{+0.019}$ & $2.43_{-0.07}^{+0.11}$\\
        & $kT_{\rm e}$ (keV) & $90_{-20}^{+35}$ & $>80$ & $>240$ & $>280$ & $>200$ & $>270$ & $60_{-30}^{+230}$\\
        \hline
        \texttt{simplcut} & $f_{\rm sc}$ & $0.121_{-0.011}^{+0.004}$ & $0.175_{-0.028}^{+0.06}$ & $0.127_{-0.023}^{+0.04}$ & $0.119_{-0.006}^{+0.007}$ & $0.114_{-0.012}^{+0.01}$ & $0.0947_{-0.008}^{+0.0025}$ & $0.076_{-0.015}^{+0.01}$\\ 
        \hline
        \texttt{cflux} \\
        $F_{\rm diskbb}$ & (10$^{-8}$~erg~cm$^{-2}$~s$^{-1}$) & $0.215_{-0.028}^{+0.03}$ & $0.60_{-0.03}^{+0.06}$ & $1.066_{-0.024}^{+0.06}$ & $1.547_{-0.029}^{+0.022}$ & $1.60_{-0.04}^{+0.05}$ & $1.838_{-0.029}^{+0.021}$ & $2.009_{-0.04}^{+0.025}$\\
        $F_{\rm relxillcp}$ & (10$^{-8}$~erg~cm$^{-2}$~s$^{-1}$) & $1.278_{-0.014}^{+0.017}$ & $1.17_{-0.11}^{+0.1}$ & $1.34_{-0.21}^{+0.05}$ & $1.21_{-0.09}^{+0.11}$ & $1.18_{-0.09}^{+0.09}$ & $1.20_{-0.09}^{+0.08}$ & $1.2_{-0.4}^{+0.7}$\\
        \hline
        & $L_{\rm 0.1-100 keV}/L_{\rm Edd}$ (\%) & 10.5 & 13.7 & 16.8 & 19.7 & 19.7 & 21.0 & 21.7 \\
        \hline
        \texttt{constant} & ME/LE & \multicolumn{7}{c}{$1.018_{-0.007}^{+0.017}$} \\
        & HE/LE & \multicolumn{7}{c}{$1.141_{-0.027}^{+0.020}$} \\ 
        \hline
                        & $\chi^2$/d.o.f &  \multicolumn{7}{c}{3734.8/3383} \\
        \hline\hline
    \end{tabular}
    \textit{Note.} Best-fit parameters for the model \texttt{tbabs*simplcut(diskbb+relxillcp)} with a broken power-law emissivity. Parameters with $^*$ are fixed during the fit and the symbol $P$ denotes the upper or lower boundary. Parameters that are only shown for Epoch~4 are tied across all observations. The model flux is before Comptonization by \texttt{simplcut}. The absorption corrected X-ray flux is calculated with the \texttt{flux} command in \texttt{XSPEC}.
\end{table*}

\begin{table*}
    \centering
    \caption{Best-fit values with \texttt{refhiden} (Model 4).}
    \label{refhiden}
    \renewcommand\arraystretch{2.0}
    \begin{tabular}{lcccccccc}
        \hline\hline
        Component       & Parameter & Epoch 1 & Epoch 2 & Epoch 3 & Epoch 4 & Epoch 5 & Epoch 6 & Epoch 7 \\
        \hline
        \texttt{tbabs} & $N_{\rm H}$ (10$^{22}$ cm$^{-2}$) & \multicolumn{7}{c}{$0.36_{-0.06}^{+0.09}$} \\
        \hline
        \texttt{relconv} & $a_*$ & \multicolumn{7}{c}{$0.998^*$} \\
        & $q_{\rm in}$ & $2.0_{-2.2}^{+2.0}$ & $2.17_{-0.16}^{+0.21}$ & $1.7_{-0.6}^{+0.5}$ & $2.35_{-0.14}^{+0.14}$ & $3.3_{-0.6}^{+0.7}$ & $2.7_{-0.3}^{+0.5}$ & $2.5_{-0.3}^{+0.8}$\\
        & $q_{\rm out}$ & \multicolumn{7}{c}{$=q_{\rm in}$} \\
        & Incl & \multicolumn{7}{c}{$33.2_{-1.4}^{+2.4}$} \\
        & $R_{\rm in}$ (ISCO) & $9_{-P}^{+P}$ & $1.5_{-P}^{+5}$ & $1.0_{-P}^{+20}$ & $1.6_{-P}^{+2.8}$ & $5.1_{-1.7}^{+2.5}$ & $3.8_{-1.8}^{+4}$ & $2.6_{-1.5}^{+2.9}$\\
        \hline
        \texttt{refhiden} & $kT_{\rm bb}$ (keV) & $0.487_{-0.008}^{+0.013}$ & $0.400_{-P}^{+0.007}$ & $0.400_{-P}^{+0.005}$ & $0.441_{-0.004}^{+0.004}$ & $0.4403_{-0.0027}^{+0.004}$ & $0.459_{-0.004}^{+0.004}$ & $0.468_{-0.004}^{+0.005}$\\
        & $H_{\rm den}~(10^{20}~{\rm cm}^{-3})$ & $>86$ & $16.2_{-1.8}^{+4.0}$ & $5.7_{-0.7}^{+0.6}$ & $5.3_{-0.5}^{+0.4}$ & $4.4_{-0.3}^{+0.5}$ & $5.4_{-0.4}^{+0.3}$ & $5.6_{-0.4}^{+0.4}$\\
        & Illum/BB & $1.8_{-0.4}^{+0.5}$ & $1.82_{-0.29}^{+0.1}$ & $1.27_{-0.2}^{+0.06}$ & $0.84_{-0.11}^{+0.09}$ & $0.97_{-0.04}^{+0.3}$ & $0.77_{-0.07}^{+0.14}$ & $0.52_{-0.06}^{+0.18}$\\
        & Norm & $0.39_{-0.03}^{+0.03}$ & $0.70_{-0.05}^{+0.06}$ & $0.89_{-0.04}^{+0.04}$ & $0.86_{-0.09}^{+0.15}$ & $0.99_{-0.09}^{+0.19}$ & $0.92_{-0.21}^{+0.07}$ & $0.73_{-0.11}^{+0.2}$\\
        \hline
        \texttt{simplcut} & $f_{\rm sc}$ & $0.773_{-0.015}^{+0.011}$ & $0.417_{-0.03}^{+0.015}$ & $0.293_{-0.017}^{+0.009}$ & $0.163_{-0.004}^{+0.005}$ & $0.132_{-0.009}^{+0.006}$ & $0.132_{-0.003}^{+0.003}$ & $0.110_{-0.011}^{+0.013}$\\
        & $\Gamma$ & $1.867_{-0.009}^{+0.013}$ & $2.124_{-0.018}^{+0.015}$ & $2.283_{-0.025}^{+0.013}$ & $2.305_{-0.016}^{+0.012}$ & $2.315_{-0.013}^{+0.026}$ & $2.381_{-0.021}^{+0.014}$ & $2.42_{-0.06}^{+0.05}$ \\
        & $kT_{\rm e}$ (keV) & $26.4_{-2.1}^{+1.6}$ & $50_{-10}^{+10}$ & $>200$ & $>100$ & $>100$ & $>360$ & $70_{-30}^{+500}$\\
        \hline
        & $L_{\rm 0.1-100 keV}/L_{\rm Edd}$ (\%) & 8.2 & 9.3 & 11.0 & 12.2 & 12.5 & 13.3 & 12.8 \\
        \hline
        \texttt{constant} & ME/LE & \multicolumn{7}{c}{$0.985_{-0.006}^{+0.008}$} \\
        & HE/LE & \multicolumn{7}{c}{$1.057_{-0.019}^{+0.016}$} \\ 
        \hline
                        & $\chi^2$/d.o.f &  \multicolumn{7}{c}{3788.8/3386} \\
        \hline\hline
    \end{tabular}
    \textit{Note.} Best-fit parameters for the model \texttt{tbabs*simplcut*relconv*refhiden} with a power-law emissivity. Parameters with $^*$ are fixed during the fit and the symbol $P$ denotes the upper or lower boundary. Parameters that are only shown for Epoch~4 are tied across all observations. The absorption corrected X-ray flux is calculated with the \texttt{flux} command in \texttt{XSPEC}.
\end{table*}

\end{document}